\documentclass[%
aip,
 jmp,%
 amsmath,amssymb,
 reprint,%
]{revtex4-1}

\usepackage{color}    
\usepackage{graphicx}
\usepackage{dcolumn}
\usepackage{mathtools}
\usepackage{bm}
\usepackage{subfigure}
\usepackage{amssymb}
\usepackage{multirow}
\usepackage{amsmath}
\usepackage{braket}

\usepackage{amssymb}

\usepackage{lineno}

\begin{document}

\title{Path Integral Monte Carlo Simulation of Degenerate Electrons:\\ Permutation-Cycle Properties}

\author{T.~Dornheim}\email{dornheim@theo-physik.uni-kiel.de}
\affiliation{Institut f\"ur Theoretische Physik und Astrophysik, Christian-Albrechts-Universit\"at zu Kiel,\\
 Leibnizstr. 15, Kiel, Germany}


\author{S.~Groth}%
 \affiliation{Institut f\"ur Theoretische Physik und Astrophysik, Christian-Albrechts-Universit\"at zu Kiel,\\
 Leibnizstr. 15, Kiel, Germany}

\author{A.V.~Filinov}

\affiliation{Institut f\"ur Theoretische Physik und Astrophysik, Christian-Albrechts-Universit\"at zu Kiel,\\
 Leibnizstr. 15, Kiel, Germany}
 
 \affiliation{Joint Institute for High Temperatures RAS, Izhorskaya Str. 13, Moscow, Russia}

\author{M.~Bonitz}
\affiliation{Institut f\"ur Theoretische Physik und Astrophysik, Christian-Albrechts-Universit\"at zu Kiel,\\
 Leibnizstr. 15, Kiel, Germany}

\date{\today}

\begin{abstract}
Being motivated by the surge of fermionic quantum Monte Carlo simulations at finite temperature, we present a detailed analysis of the permutation-cycle properties of path integral Monte Carlo (PIMC) simulations of degenerate electrons. Particular emphasis is put onto the uniform electron gas in the warm dense matter regime. We carry out PIMC simulations of up to $N=100$ electrons and investigate exchange-cycle frequencies, which are found not to follow any simple exponential law even in the case of ideal fermions due to the finite size of the simulation box. Moreover, we introduce a permutation-cycle correlation function, which allows us to analyse the joint probability to simultaneously find cycles of different lengths within a single configuration. Again, we find that finite-size effects predominate the observed behaviour. Finally, we briefly consider an inhomogeneous system, namely electrons in a $2D$ harmonic trap. We expect our results to be of interest for the further development of fermionic PIMC methods, in particular to alleviate the notorious fermion sign problem.
\end{abstract}

\pacs{Valid PACS appear here}
\maketitle

\section{Introduction}

The well-known path-integral Monte-Carlo (PIMC) method~\cite{berne_JCP_1982,pollock_PRB_1984,ceperley_path_1995}
constitutes a highly successful tool for the simulation of distinguishable particles (often referred to as \emph{boltzmannons}) and bosons. Due to its exact nature, PIMC has been pivotal for the understanding of important physical effects such as superfluidity~\cite{sindzingre_path-integral_1989,kwon_local_2006,dornheim_superfluidity_2015} and Bose-Einstein condensation~\cite{pilati_dilute_2010,noauthor_path-integral_2016} and, using state-of-the-art
Monte-Carlo sampling techniques like the worm algorithm by Boninsegni \textit{et al.}~\cite{boninsegni_worm_2006-1,boninsegni_worm_2006}, simulations of up to $N\sim10^4$ particles are possible.

In stark contrast, PIMC simulations of fermions like He$^3$ or electrons are severely limited by the notorious fermion sign problem~\cite{loh_sign,doi:10.1063/1.4977920}. This is a direct consequence of the anti-symmetry under particle-exchange, leading to a cancellation of positive and negative terms, which can result in an almost vanishing signal-to-noise ratio. Consequently, PIMC simulations of electrons are restricted to relatively high temperature or strong coupling, but break down when quantum-degeneracy effects become important. This is very unfortunate, as fermionic quantum Monte-Carlo simulations at finite temperature are highly needed for the description of, e.g., ultracold atoms~\cite{RevModPhys.80.885,PhysRevLett.74.2288,Bernu1992}, lattice models~\cite{PhysRevB.72.035122,0295-5075-90-1-10004,Cheuk1260}, or even exotic quark-gluon plasmas~\cite{PhysRevC.87.035207,doi:10.1002/ctpp.201400056}.

An application of particular interest is so-called \emph{warm dense matter}---an extreme state occurring in astrophysical objects like giant planet interiors~\cite{PhysRevLett.120.115703,PhysRevLett.120.025701,PhysRevB.75.024206}, on the pathway towards inertial confinement fusion~\cite{PhysRevB.84.224109,PhysRevE.90.033111}, or in state-of-the art experiments with, e.g., free-electron lasers or diamond anvil cells, see Ref.~\onlinecite{falk_2018} for a topical review. More specifically, the warm dense matter regime is characterized by two parameters that are both of the order of one: 1) the density parameter (or coupling parameter, \emph{Wigner-Seitz} radius\cite{seitz1943physics}) $r_s=\overline{r}/a_\textnormal{B}$, with $\overline{r}$  and $a_\textnormal{B}$ being the average inter-particle distance and Bohr radius, and 2) the degeneracy temperature $\theta=k_\textnormal{B}T/E_\textnormal{F}$, where $E_\textnormal{F}$ denotes the usual Fermi energy~\cite{giuliani2005quantum}. Therefore, the intricate and nontrivial interplay of a) Coulomb coupling, b) thermal excitation, and c) quantum degeneracy effects renders a thorough theoretical description most challenging~\cite{graziani2014frontiers}, leaving \textit{ab initio} quantum Monte Carlo methods as the option of choice\cite{10.1007/978-3-319-04912-0_5}.
Consequently, over the last years, there has been a remarkable spark of new developments in the field of thermodynamic quantum Monte Carlo simulations of electrons at finite temperature~\cite{dornheim_review,dornheim_NJP2015,doi:10.1063/1.4936145,PhysRevLett.117.156403,PhysRevLett.119.135001,PhysRevB.89.245124,doi:10.1063/1.4927434,PhysRevLett.117.115701,PhysRevLett.115.050603,doi:10.1021/acs.jctc.8b00569,PhysRevB.95.205109,PhysRevLett.115.176403,PhysRevE.91.033108,universe4120133,PhysRevE.96.023203,doi:10.1063/1.4999907,PhysRevLett.121.255001}.
Yet, despite this impressive progress, there still does not exist a single method that is capable to provide an accurate description of a correlated Fermi system for all parameters~\cite{doi:10.1063/1.4977920}, which makes the further development of state-of-the-art QMC simulations indispensable.

In this article, we present a detailed analysis of the application of the path-integral Monte-Carlo method to a system of correlated electrons at finite temperature. More specifically, we investigate the permutation properties of the uniform electron gas in the warm dense matter regime~\cite{PhysRevLett.110.146405,PhysRevB.93.085102,PhysRevB.93.205134,doi:10.1002/ctpp.201700096,dornheim_review}, an important model system, which has been fully described only recently~\cite{dornheim_review,PhysRevLett.119.135001,doi:10.1002/ctpp.201600082}.
In fact, the particular structure of the permutation space within a path-integral Monte-Carlo calculation fully determines the degree of severity of the fermion sign problem, and, therefore, whether simulations are feasible~\cite{PhysRevA.48.4075,PhysRevE.80.066702}. In particular, if permutations of particles were independent of each other, the full fermionic configuration space could be significantly simplified as has been recently proposed in Ref.~\onlinecite{dubois_permutation}, and PIMC simulations were possible at parameters that are currently out of reach for other methods. The main goal of the present work is to investigate correlation effects (which includes, but is not limited to effects caused by the Coulomb coupling between electrons) within the permutation cycles and, in this way, to assess if further simplifications are possible.

The paper is organised as follows: in Sec.~\ref{sec:theory}, we introduce the theoretical background of the PIMC method (\ref{sec:PIMC}), how it needs to be adapted to the simulation of identical particles (\ref{sec:ident}), and, in the case of fermions, how this leads to the fermion sign problem (\ref{sec:FSP}). Furthermore, in Sec.~\ref{sec:permutation_cycle_theory} we introduce several quantities to measure the desired permutation cycle properties and give a few useful formulas for the noninteracting (ideal) case (\ref{sec:ideal_theory}). In Sec.~\ref{sec:results}, we discuss our simulation results starting with noninteracting fermions (\ref{sec:results_ideal}), which, remarkably, exhibit a quite nontrivial behaviour due to the finite size of the simulation box. The bulk of our results is devoted to the warm dense electron gas (\ref{sec:ueg_results}), where we investigate different system sizes, coupling strengths, and temperatures. Further, in Sec.~\ref{sec:itsatrap} we extend our considerations to electrons in a $2D$ harmonic trap, where the finite-size effects are an inherent feature of the system.
Lastly, in Sec.~\ref{sec:discusson} we summarise our findings and discuss the implications on the future development of PIMC simulations of electrons at finite temperature.


\section{Theory\label{sec:theory}}

\subsection{Path Integral Monte Carlo\label{sec:PIMC}}

Let us consider a system of $N$ distinguishable particles in a three-dimensional box of length $L$ and volume $V=L^3$ at an inverse temperature $\beta=1/k_\textnormal{B}T$. In thermodynamic equilibrium, such a system is fully described by the canonical partition function
\begin{eqnarray}\label{eq:trace}
Z = \textnormal{Tr}\ \hat\rho = \textnormal{Tr}\ e^{-\beta\hat H} \quad ,
\end{eqnarray}
with the Hamiltonian $\hat H = \hat V + \hat K$ being the sum of a potential and a kinetic contribution, respectively.
In coordinate space, Eq.~(\ref{eq:trace}) becomes
\begin{eqnarray}\label{eq:trace2}
Z = \int \textnormal{d}\mathbf{R}\ \bra{\mathbf{R}} e^{-\beta\hat H} \ket{\mathbf{R}}
= \int \textnormal{d}\mathbf{R}\ \rho(\mathbf{R},\mathbf{R},\beta)
\quad ,
\end{eqnarray}
with $\mathbf{R}=(\mathbf{r}_1,\dots,\mathbf{r}_N)^T$ containing the coordinates of all $N$ particles, and $\rho(\mathbf{R},\mathbf{R}',\beta)$ being the thermal density matrix. Unfortunately, Eq.~(\ref{eq:trace2}) is not directly useful since the matrix elements of $\hat\rho$ cannot be readily evaluated as $\hat V$ and $\hat K$ do not commute,
\begin{eqnarray}\label{eq:factorization_error}
e^{-\beta(\hat V + \hat K)} \neq e^{-\beta \hat V} e^{-\beta \hat K} \quad .
\end{eqnarray}
To overcome this obstacle, we use a semi-group property of the density operator,
\begin{eqnarray}
e^{-\beta \hat H} = \prod_{\alpha=0}^{P-1} e^{-\epsilon\hat H} \quad ,
\end{eqnarray}
 with $\epsilon = \beta / P$, and, according to the well-known
Trotter formula~\cite{trotter_product_1959,de_raedt_applications_1983},
the factorization error from Eq.~(\ref{eq:factorization_error}) vanishes in the limit of large $P$,
\begin{eqnarray}
e^{-\beta(\hat V + \hat K)} = \lim_{P\to\infty} \left(
e^{-\epsilon\hat V} e^{-\epsilon \hat K} 
\right)^P\quad .
\end{eqnarray}
In a nutshell, we can express the density-matrix elements at some temperature $T$, $\rho(\mathbf{R},\mathbf{R}',\beta)$, as a product of $P$ density-matrices, but at a $P$-times higher temperature, $\rho(\mathbf{R},\mathbf{R}',\epsilon)$. This is advantageous, since, for sufficiently large $P$, we can evaluate $\rho(\mathbf{R},\mathbf{R}',\epsilon)$ using a suitable high-temperature approximation, with $P$ being a convergence parameter within the PIMC formalism.
The partition function thus becomes
\begin{eqnarray}\label{eq:Z2}
Z = \int \textnormal{d}\mathbf{R}\textnormal{d}\mathbf{R}_1\dots\textnormal{d}\mathbf{R}_{P-1} \prod_{\alpha=0}^{P-1}\rho(\mathbf{R}_i,\mathbf{R}_{i+1},\epsilon) \quad ,
\end{eqnarray}
with $\mathbf{R} = \mathbf{R}_0 = \mathbf{R}_P$ due to the definition of the trace. 
Formally, the density matrices in Eq.~(\ref{eq:Z2}) are equivalent to propagators in the imaginary time
$\tau\in[0,\beta]$ by a time step $\epsilon$, and the trace is then interpreted as the sum over all closed paths $\mathbf{X}$.\begin{figure*}
\centering\hspace*{-0.6cm}
\includegraphics[width=0.37\textwidth]{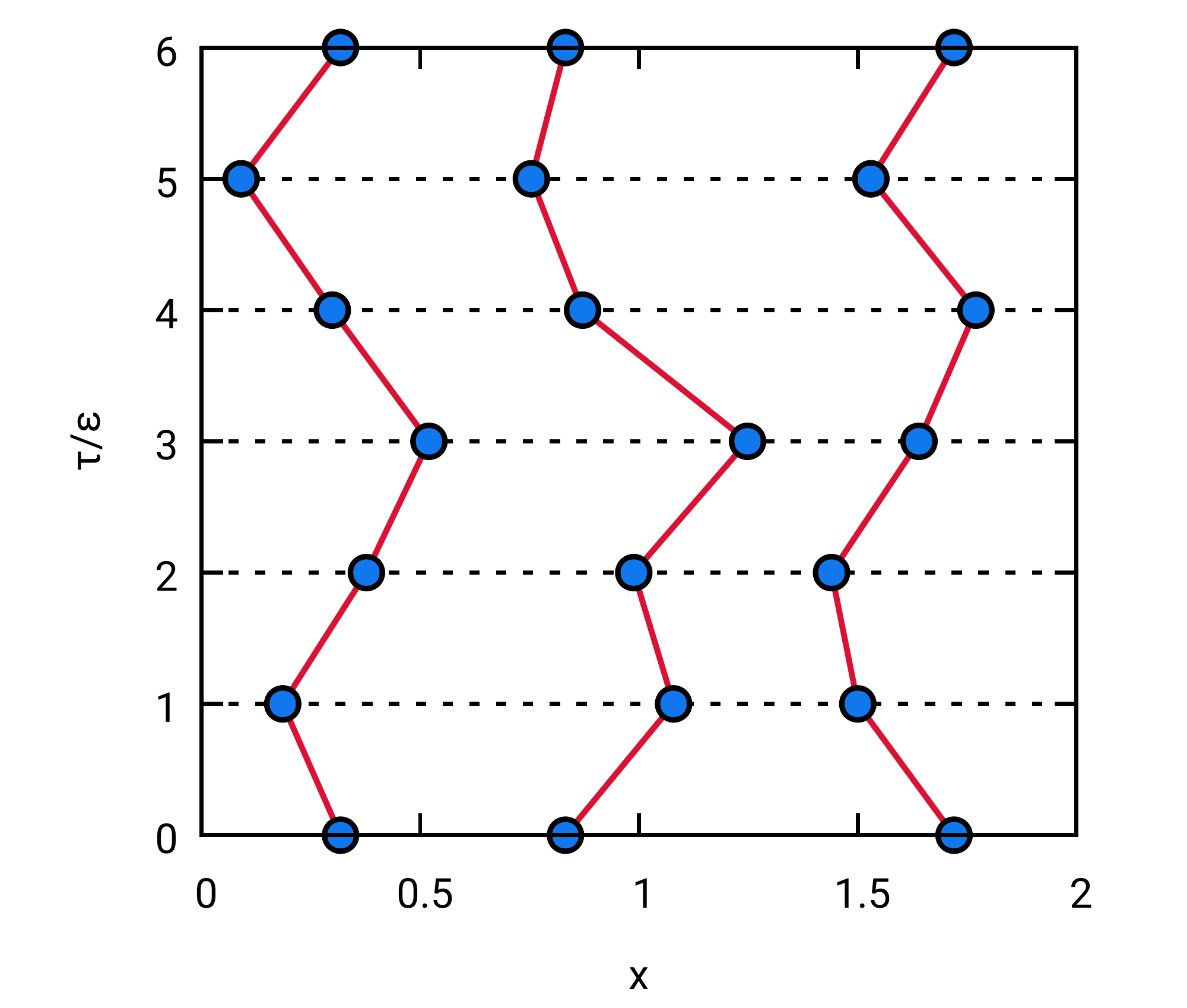}\hspace*{-0.6cm}
\includegraphics[width=0.37\textwidth]{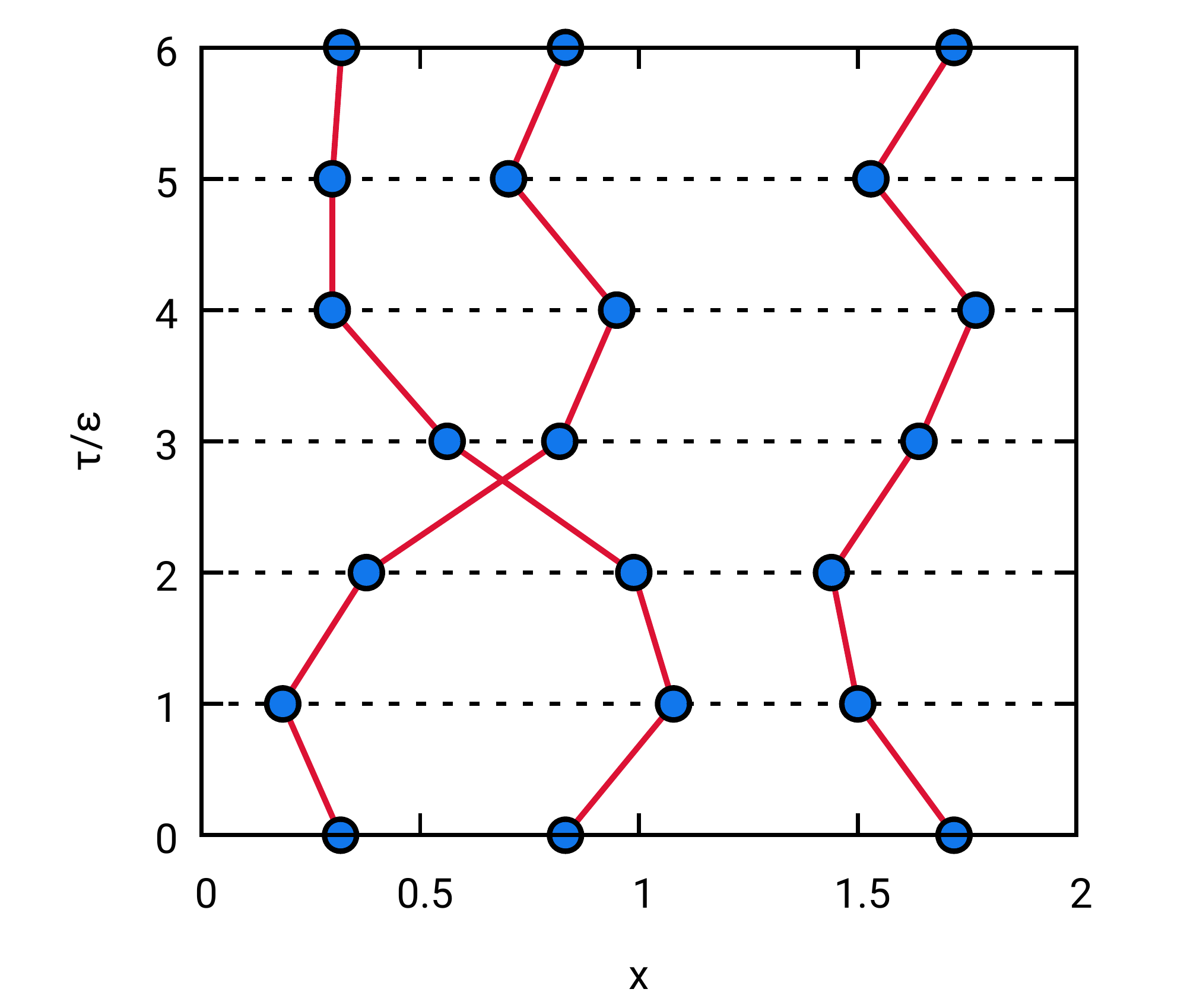}\hspace*{-0.6cm}
\includegraphics[width=0.37\textwidth]{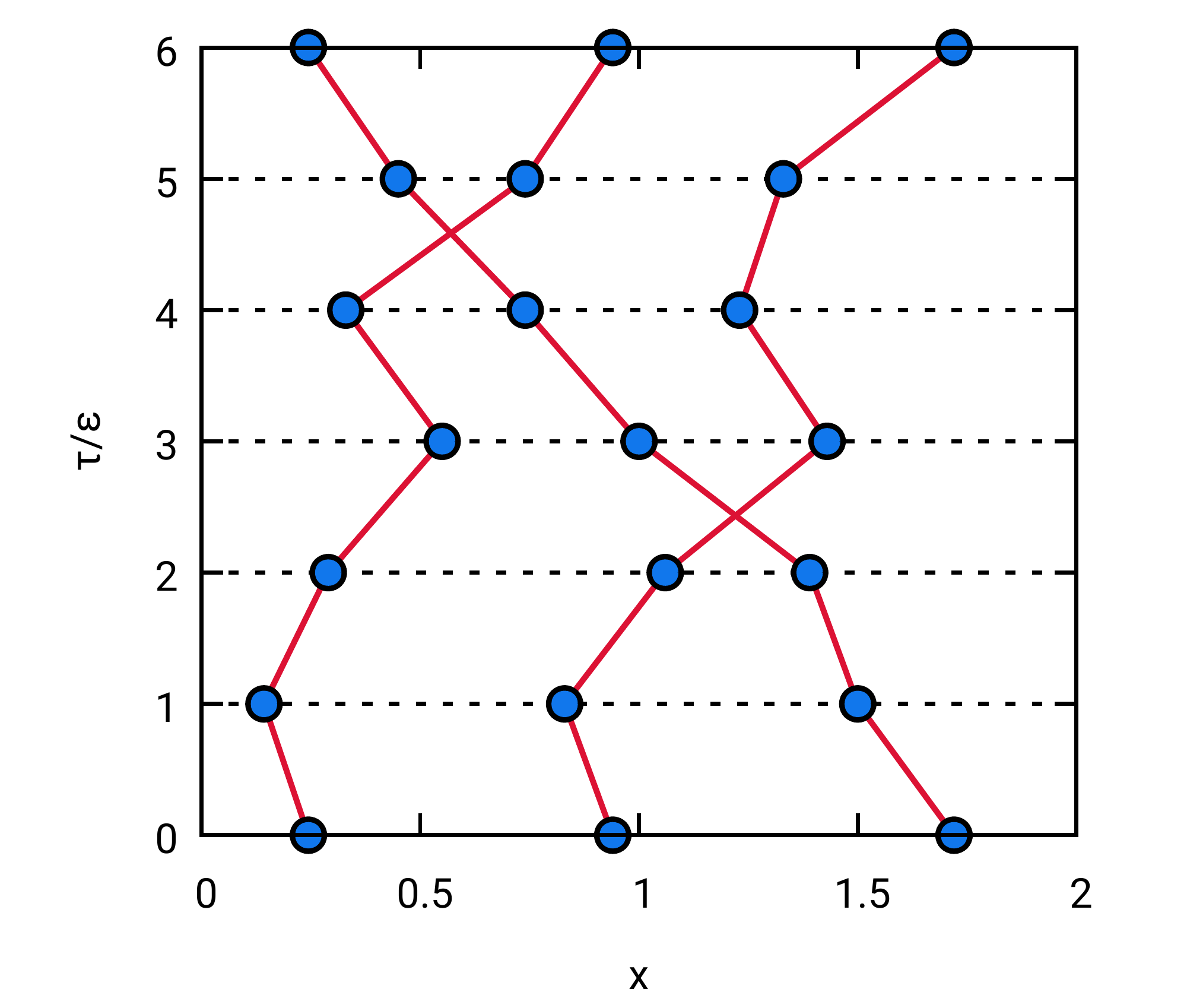}
\caption{\label{fig:illustration}
Possible permutation classes of three spin-polarized electrons in the path-integral Monte-Carlo formalism---Shown are configurations in the $x$-$\tau$-plane with no exchange (left), one pair-exchange and a single particle (center), and two pair-exchanges leading to one three-particle cycle (right). The corresponding fermionic configuration weights $W_\textnormal{F}(\mathbf{X})$ are positive, negative, and positive, respectively.
}
\end{figure*} This is illustrated in the left panel of Fig.~\ref{fig:illustration}, where we show a configuration of $N=3$ particles---each being represented by an entire closed path of $P=6$ sets of coordinates---in the $x$-$\tau$-plane.
In this way, the complicated quantum mechanical system of interest has effectively been mapped onto a system of classical ring-polymers~\cite{chandler_exploiting_1981}.
The only seeming drawback of this procedure is the drastic increase in the dimensionality in Eq.~(\ref{eq:Z2}), which makes the application of standard quadrature schemes unfeasible. Fortunately, this obstacle can be overcome by the utilization of stochastic, i.e., Monte-Carlo methods, which are not afflicted by this curse of dimensionality~\cite{yellow_book,liu2013monte}.
In particular, the basic idea of the path-integral Monte-Carlo method~\cite{berne_JCP_1982,pollock_PRB_1984,ceperley_path_1995} is to use the Metropolis algorithm~\cite{metropolis_equation_1953} to randomly generate a Markov chain of $N_\textnormal{MC}$ particle configurations $\mathbf{X}$, $\{\mathbf{X}\}_\textnormal{MC}$, distributed according to their respective contribution $W(\mathbf{X})$ to the partition function,
\begin{eqnarray}
Z = \int \textnormal{d}\mathbf{X}\ W(\mathbf{X}) \quad ,
\end{eqnarray}
with $\textnormal{d}\mathbf{X}=\textnormal{d}\mathbf{R}_0\dots\textnormal{d}\mathbf{R}_{P-1}$.
The expectation value of a given observable $\hat A$ is then computed as the sum over these elements,
\begin{eqnarray}\label{eq:MC}
\braket{\hat A} \approx \frac{1}{N_\textnormal{MC} } \sum_{\mathbf{X}\in\{\mathbf{X}\}_{\textnormal{MC}}} A(\mathbf{X}) \quad ,
\end{eqnarray}
with $A(\mathbf{X})$ being the so-called estimator. For completeness, we note that for any finite number of Monte-Carlo samples $N_\textnormal{MC}$, Eq.~(\ref{eq:MC}) is subject to a statistical uncertainty $\Delta A$, which vanishes as
\begin{eqnarray}
\Delta A \sim \frac{1}{\sqrt{N_\textnormal{MC}}} \quad .
\end{eqnarray}
Hence, the Monte-Carlo error can---at least in principle---be made arbitrarily small and the PIMC approach is \emph{quasi-exact}.

A more detailed introduction of the PIMC method including a description of sampling-schemes and the computation of observables is beyond the scope of the present work and the interested reader is referred to, e.g., Refs.~\onlinecite{ceperley_path_1995,boninsegni_worm_2006-1,dornheim_review}.

\subsubsection{Identical Particles: PIMC simulation of bosons and fermions\label{sec:ident}}

Let us next extend our consideration to the simulation of $N$ spin-polarized bosons or fermions. The partition function is then given by~\cite{ceperley_path_1995}
\begin{eqnarray}\label{eq:Z_fermi}
Z_{\textnormal{B,F}} =  \frac{1}{N!} \sum_{\sigma\in S_N} \int \textnormal{d}&\mathbf{X}&\
\left( \pm 1 \right)^{T} 
\rho(\mathbf{R}_0,\hat\pi_\sigma\mathbf{R}_{1},\epsilon)\\ \nonumber & & \prod_{\alpha=1}^{P-1}\rho(\mathbf{R}_i,\mathbf{R}_{i+1},\epsilon) \quad ,
\end{eqnarray}
and incorporates a summation over all elements $\sigma$ of the permutation group $S_N$, with $\hat \pi_\sigma$ being the corresponding permutation operator. The plus and minus sign corresponds to bosons and fermions, respectively, which means that the sign of the configuration weight function $W(\mathbf{X})$ alternates with the number of pair permutations $T$ in the latter case. This renders fermionic PIMC simulations somewhat tricky, as we shall see in Sec.~\ref{sec:FSP}.

In practice, Eq.~(\ref{eq:Z_fermi}) implies that we now also have to stochastically generate configurations with trajectories where the start- and end-points of a given particle are not identical. This leads to so-called exchange-cycles, which are illustrated in Fig.~\ref{fig:illustration} for a system of $N=3$ identical particles. The left panel shows a configuration without any permutations, such as we have already encountered in the discussion of PIMC simulations of distinguishable particles. Consequently, the corresponding fermionic configuration weight $W_\textnormal{F}(\mathbf{X})$ is positive. In contrast, the central panel depicts a configuration with a single pair-exchange, leading to one path containing two particles and one path containing only one. In the case of fermions, the sign of the weight function is negative, $W_\textnormal{F}(\mathbf{X})\leq 0$. Finally, in the right panel we show a configuration with two pair-permutations, leading to only a single path, which contains all $N=3$ particles with a positive configuration weight.

As a side remark, we mention that the efficient realization of such exchange-cycles was solved only relatively recently by the worm algorithm introduced in Refs.~\onlinecite{boninsegni_worm_2006-1,boninsegni_worm_2006}. All results shown in the present work have been obtained using a canonical adaption of this approach.

In the case of bosons, such as ultracold atoms~\cite{PhysRevLett.96.105301,PhysRevLett.97.045301,PhysRevLett.105.070401} or composite particles like indirect excitons~\cite{PhysRevB.84.075130}, all contributions to $Z$ in Eq.~(\ref{eq:Z_fermi}) are strictly positive and PIMC simulations have been successfully used to study important phenomena such as superfluidity~\cite{sindzingre_path-integral_1989,kwon_local_2006,dornheim_superfluidity_2015}, Bose-Einstein condensation~\cite{pilati_dilute_2010,noauthor_path-integral_2016}, and collective excitations~\cite{filinov_collective_2012,filinov_correlation_2016}. Unfortunately, for fermions, such as the ubiquitous electrons, this does not hold true, and PIMC simulations are afflicted with the notorious fermion sign problem~\cite{loh_sign,doi:10.1063/1.4977920}.

\subsubsection{The fermion sign problem\label{sec:FSP}}

As we have seen in the previous section, a direct utilization of the Metropolis algorithm to generate a Markov chain of fermionic configurations $\mathbf{X}$ distributed according to their respective weight function $W_\textnormal{F}(\mathbf{X})$ is not possible, as probabilities must not be negative. To work around this issue, we can generate a Markov chain of configurations $\mathbf{X}$ distributed according to the absolute value of the fermionic weight function
\begin{eqnarray}\label{eq:Z_modified}
Z' = \int \textnormal{d}\mathbf{X}\ |W_\textnormal{F}(\mathbf{X})| = Z_\textnormal{B} \quad ,
\end{eqnarray}
which, in the case of standard PIMC, are nothing else than the bosonic weights,
as $|W_\textnormal{F}(\mathbf{X})| = W_\textnormal{B}(\mathbf{X})$.
The fermionic expectation value of an observable $\hat A$ is then computed as 
\begin{eqnarray}\label{eq:AS}
\braket{\hat A} = \frac{ \braket{\hat A \hat S}' }{\braket{\hat S}'} \quad ,
\end{eqnarray}
with $S(\mathbf{X})= W_\textnormal{F}(\mathbf{X}) / W_\textnormal{B}(\mathbf{X})$ being the sign of a particular configuration $\mathbf{X}$, and the notation $\braket{\dots}'$ referring to the expectation value with respect to the bosonic weights
\begin{eqnarray}
\braket{\hat A}' = \frac{1}{Z_\textnormal{B}}\int \textnormal{d}\mathbf{X}\ W_\textnormal{B}(\mathbf{X}) A(\mathbf{\mathbf{X}})  = \braket{\hat A}_\textnormal{B} \quad .
\end{eqnarray}
The denominator in Eq.~(\ref{eq:AS}) is the so-called average sign
\begin{eqnarray}
S \coloneqq \braket{\hat S}' &=& \frac{1}{Z_\textnormal{B}} \int  \textnormal{d}\mathbf{X}\ W_\textnormal{B}(\mathbf{X}) S(\mathbf{X})\\ \nonumber &=& \frac{Z_\textnormal{F}}{Z_\textnormal{B}} = e^{-\beta N (f_\textnormal{F}-f_\textnormal{B})} \quad ,
\end{eqnarray}
which is readily identified as the ratio of the fermionic and bosonic partition function (with $f$ denoting the free energy per particle), and constitutes a measure for the amount cancellation of positive and negative contributions within the PIMC simulation~\cite{doi:10.1002/ctpp.201800157,doi:10.1063/1.4977920,dornheim_review}.
In particular, the statistical uncertainty of Eq.~(\ref{eq:AS}) is inversely proportional to $S$,
\begin{eqnarray}\label{eq:fsp_error}
\Delta A_\textnormal{F} \sim \frac{1}{S \sqrt{N_\textnormal{MC}}} \sim \frac{e^{\beta N (f_\textnormal{F}-f_\textnormal{B})}}{\sqrt{N_\textnormal{MC}}} \quad ,
\end{eqnarray}
and thus exponentially increases both with system size $N$ and towards low temperature. This can only be compensated by increasing the number of Monte Carlo samples as $\sim 1 / \sqrt{N_\textnormal{MC}}$, which quickly becomes unfeasible as one runs into an \emph{exponential wall}. This is the origin of the notorious fermion sign problem~\cite{loh_sign}, which limits standard PIMC simulations of electrons to relatively high temperature or strong coupling and has been shown to be $NP$-hard for a certain class of Hamiltonians~\cite{troyer_sign}.


\subsection{Permutation Cycle Properties\label{sec:permutation_cycle_theory}}

\begin{figure}
\centering
\includegraphics[width=0.364\textwidth]{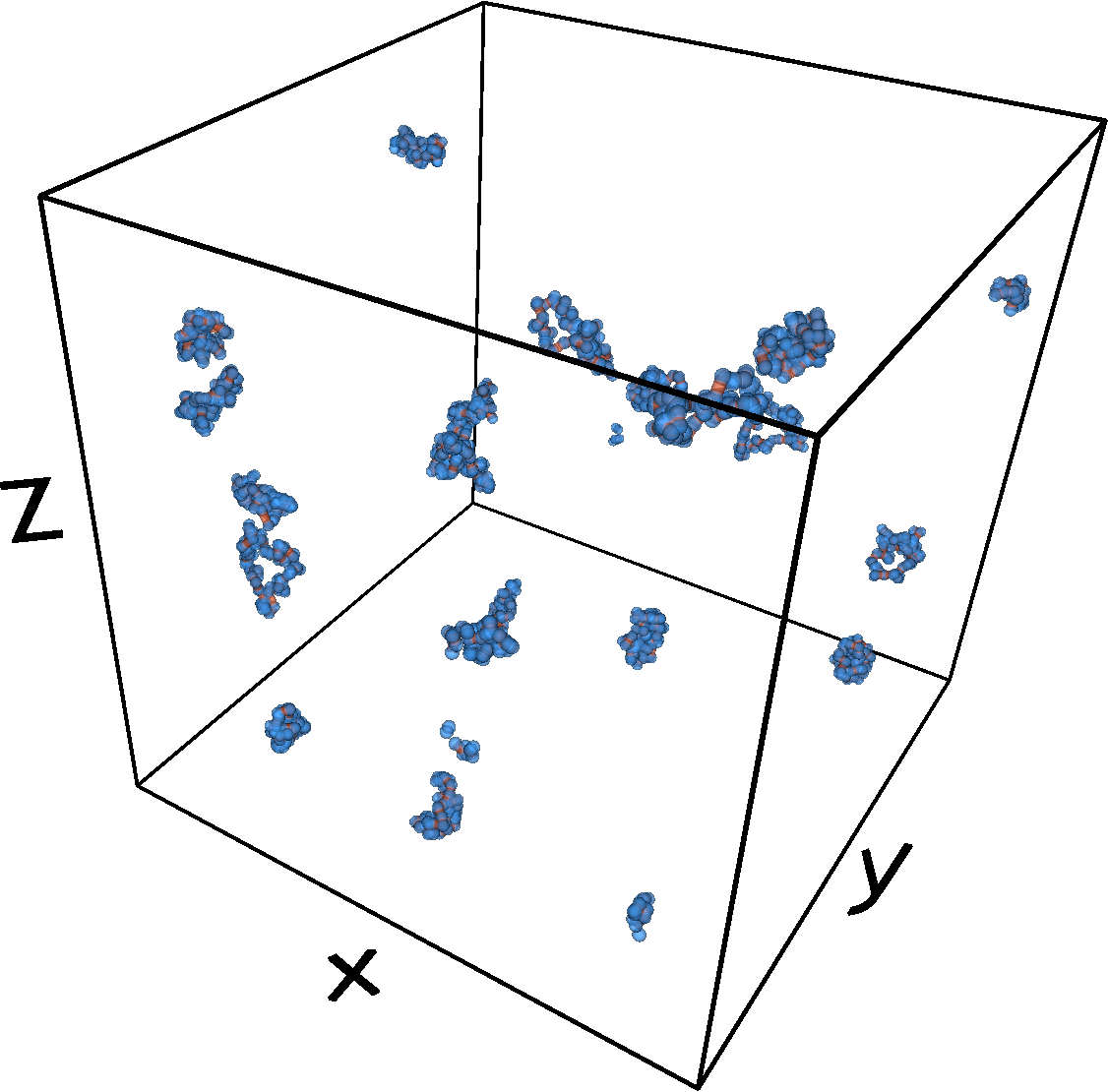}\\
\vspace*{0.5cm}
\includegraphics[width=0.364\textwidth]{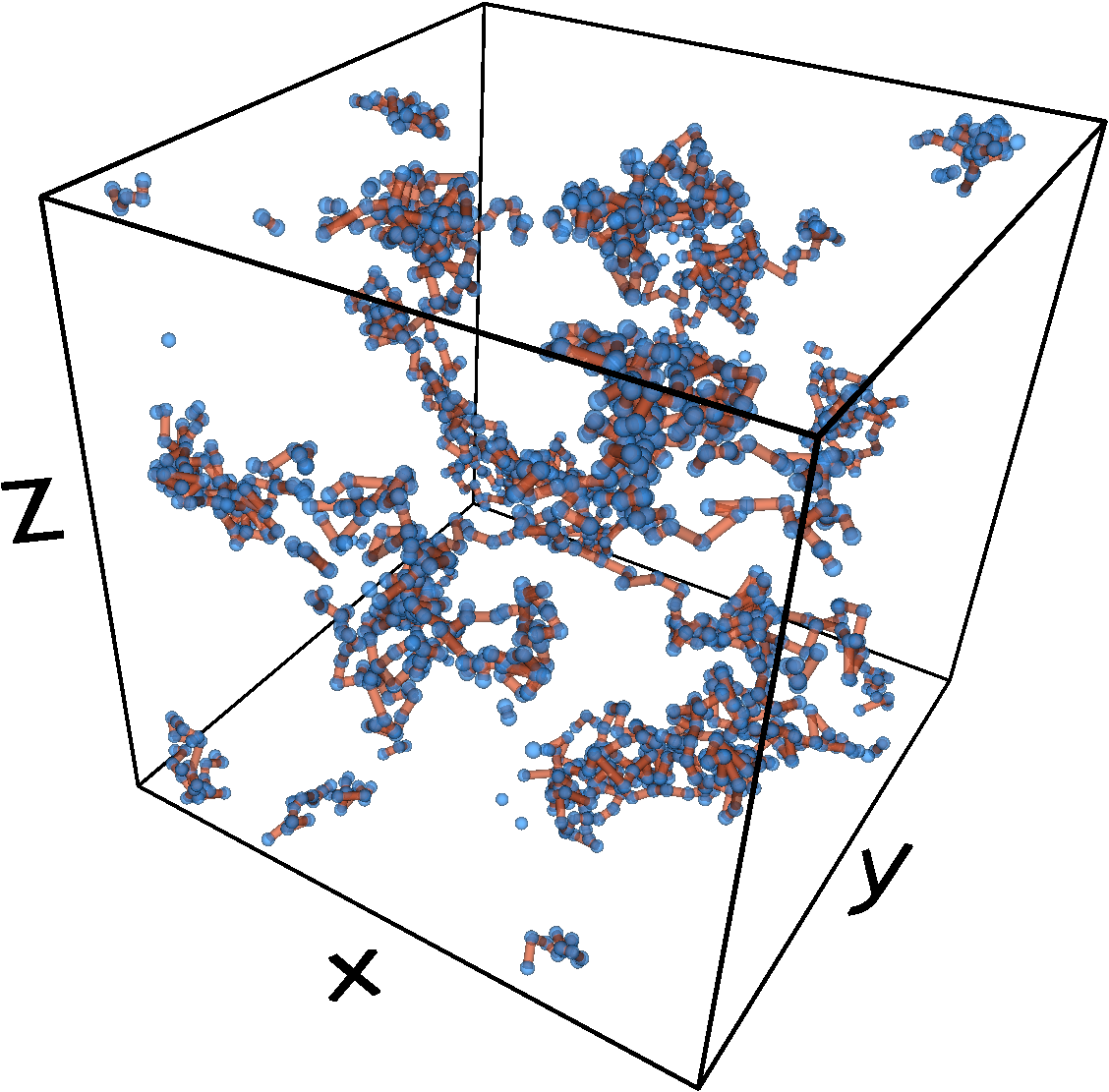}
\caption{\label{fig:snapshots}
Snapshots from PIMC simulations of the uniform electron gas at metallic density, $r_s=2$, with $N=19$ and $P=100$ for $\theta=4$ (top) and $\theta=0.5$ (bottom).
}
\end{figure}

Evidently, the feasibility of a standard PIMC simulation of electrons depends on the probability of pair-exchanges or, more specifically, on the prevalence of different exchange-cycles. This is illustrated in Fig.~\ref{fig:snapshots}, where we show snapshots from a PIMC simulation of $N=19$ spin-polarized electrons with $P=100$ high-temperature factors at a metallic density, $r_s=2$. The top panel corresponds to a relatively high temperature, $\theta=4$. In this case, the extension of the paths of different particles, which is proportional to the thermal wavelength $\lambda_\beta=\sqrt{\hbar^2 2\pi \beta / m}$, is significantly smaller than the average inter-particle distance $\overline{r}$. Consequently, fermionic exchange-effects are not of paramount importance, and exchange-cycles only seldom occur within a PIMC simulation, which results in an average sign of $S\approx0.7$. In stark contrast, the bottom panel depicts a snapshot for $\theta=0.5$, which falls into the important warm dense matter regime.
In this case, $\lambda_\beta$ is comparable to $\overline{r}$, and fermionic exchange-effects predominate. In fact, as we shall see in Sec.~\ref{sec:results}, at these conditions even \textit{macroscopic} exchange-cycles containing all $N$ particles have a significant weight. 
Accordingly, configurations with positive and negative configuration weights occur with a similar frequency, the average sign vanishes within the given statistical uncertainty [$S=0.0004(3)$], and the resulting cancellation renders standard PIMC simulations unfeasible in this regime.

In the following section, we will introduce two quantities that allow for a more rigorous characterization of the permutation cycle properties of a PIMC simulation.

The probability to find a permutation cycle of length $l$ can be readily defined as
\begin{eqnarray}\label{eq:defintion_pl}
P(l) = \frac{1}{lN}  \left< \sum_{i=1}^{N} \delta(i,l) \right> \quad ,
\end{eqnarray}
with $\delta(i,l)$ vanishing, except when particle $i$ is involved in an exchange-cycle of length $l$. Note that the pre-factor $1/l$ makes us count each cycle of length $l$ only once in the definition of $P(l)$. The PIMC expectation value for Eq.~(\ref{eq:defintion_pl}) is then expressed as 
\begin{eqnarray}\label{eq:p}
P(l) = \frac{1}{N_\textnormal{MC}} \sum_{\mathbf{X}\in\{\mathbf{X}\}_\textnormal{MC}} \left(
\frac{1}{N} \sum_{m\in{\{m\}_\mathbf{X}}} \delta_{m,l}
\right) \quad ,
\end{eqnarray}
with the second sum being carried out over all lengths $m$ in the set of permutation cycles $\{m\}_\mathbf{X}$ present in a particular Monte-Carlo configuration $\mathbf{X}$. Note that, even in the case of a fermionic PIMC simulation, Eq.~(\ref{eq:p}) is always computed as a bosonic expectation value, i.e., an observable in the modified configuration space $Z'=Z_\textnormal{B}$.

For the ideal (i.e., non-interacting) system, Eq.~(\ref{eq:defintion_pl}) can be computed semi-analytically from different partition functions (see Sec~\ref{sec:ideal_theory}) as~\cite{krauth_book_2006statistical}
\begin{eqnarray}\label{eq:analytical_pl}
P(l) = \frac{ Z_{\textnormal{B},1}(l\beta) Z_{\textnormal{B},N-l}(\beta) }{ l\ Z_{\textnormal{B},N(\beta) }} \quad ,
\end{eqnarray}
which can be used to check the validity of our implementation.


One of the central questions to be investigated in this work is whether the probability to find a cycle of length $l$ within a PIMC simulation depends on the presence of other permutation cycles or, more specifically, on $\{m\}_\mathbf{X}$. To answer this question, we define a permutation-cycle correlation function 
\begin{eqnarray}
P(l,k) = \frac{2}{l k N (N-1)}  \left< \sum_{i=1}^{N}\sum_{j>i}^N \delta(i,l)\delta(j,k) \right>\ ,
\end{eqnarray}
which is evaluated in our PIMC simulations as 
\begin{eqnarray}\label{eq:pcf}
P(l,k) = & &\frac{1}{N_\textnormal{MC}} \sum_{\mathbf{X}\in\{\mathbf{X}\}_\textnormal{MC}} \left(
\frac{1}{N(N-1)}\right. \\ \nonumber & & \left.
\sum_{m\in{\{m\}_\mathbf{X}}}\sum_{n\in{\{n\}_\mathbf{X}}} \delta_{m,l}\delta_{n,k}(1-\delta_{m,n})
\right) \quad .
\end{eqnarray}
In particular, if the probability for a cycle of length $l$ is independent of the presence of other cycles, Eq.~(\ref{eq:pcf}) should coincide with the uncorrelated quantity
\begin{eqnarray}\label{eq:pcf_uncorrelated}
P_\textnormal{u}(l,k) = P(l) P(k) \quad ,
\end{eqnarray}
for all $l$ and $k$.

\subsection{Ideal Bose and Fermi gas\label{sec:ideal_theory}}

To check the validity of our implementation and analyze our PIMC results for correlated electrons, it is useful to consider the ideal system where $P(l)$ can be computed from Eq.~(\ref{eq:analytical_pl}).
The ideal canonical partition function obeys a recursion relation with respect to the system size $N$ of the form~\cite{krauth_book_2006statistical}
\begin{eqnarray}\label{eq:Z_ideal_recursion}
Z_{\textnormal{B},N}(\beta) = \frac{1}{N} \sum_{\eta=1}^N Z_{\textnormal{B},1}(\eta\beta) Z_{\textnormal{B},N-\eta}(\beta) \quad ,
\end{eqnarray}
with $Z_{\textnormal{B},0}=1$. The initial condition for Eq.~(\ref{eq:Z_ideal_recursion}) is given by the single particle partition function $Z_{\textnormal{B},1}$, which for a three-dimensional periodic box of length $L$ is given by 
\begin{eqnarray}\label{eq:Z_single}
Z_{\textnormal{B},1}(\beta) &=& \left( \sum_{x=-\infty}^\infty e^{-\beta E_x} \right)^3\ , \\ \nonumber  E_x &=& \left( \frac{2\pi x}{L} \right)^2 \frac{1}{2} \quad ,
\end{eqnarray}
and has to be evaluated at different multiples of the inverse temperature $\beta$.
In the limit of a large box (compared to the thermal wavelength, $L \gg \lambda_\beta$), the sum in Eq.~(\ref{eq:Z_single}) can be transformed into a continuous integral, and we obtain the simple expression
\begin{eqnarray}\label{eq:Z_approximate}
\sum_{x} \to \int \textnormal{d}x \Rightarrow Z_{\textnormal{B},1}(\beta) = \frac{V}{\lambda_\beta^3} \quad .
\end{eqnarray}

\begin{figure}
\centering
\includegraphics[width=0.4\textwidth]{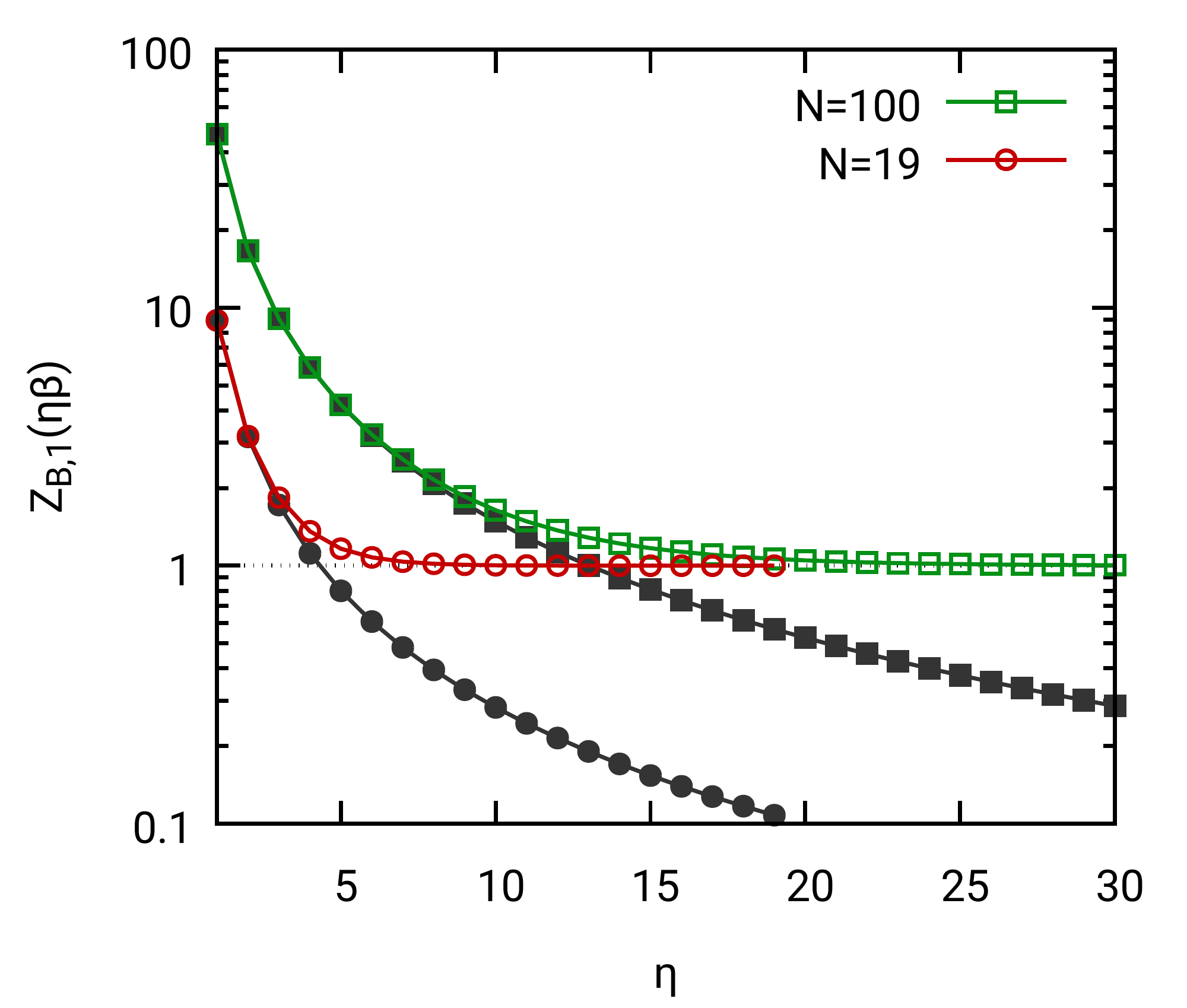}
\caption{\label{fig:Z_single}
Partition function $Z_{\textnormal{B},1}$ for a single particle in a periodic box corresponding to $\theta=0.5$ and $r_s=2$ for $N=100$ (squares) and $N=19$ (boxes). The colored and dark grey symbols correspond to the exact infinite sum [Eq.~(\ref{eq:Z_single})] and approximate analytical result [Eq.~(\ref{eq:Z_approximate})], respectively.
}
\end{figure}

In Fig.~\ref{fig:Z_single}, we show $Z_{\textnormal{B},1}(\eta \beta)$ computed from Eq.~(\ref{eq:Z_single}) for the relevant multiples of $\beta$ for the UEG with $N=100$ (green squares) and $N=19$ (red circles) at $\theta=0.5$. With decreasing temperature (i.e., increasing inverse temperature $\eta\beta$), only the lowest state in Eq.~(\ref{eq:Z_single}) is occupied, and $Z$ converges to one.
In addition, the dark grey symbols depict the results for $Z_{\textnormal{B},1}(\eta \beta)$ from the continuous approximation Eq.~(\ref{eq:Z_approximate}). At large temperature (small $\eta$), replacing the discrete sum by an integral is accurate, but with increasing $\eta$ the thermal wavelength eventually becomes comparable to $L$ and the approximation breaks down. Naturally, this happens at slightly lower temperature for the larger system.
Thus, we conclude that a numerical evaluation of the infinite sum, Eq.~(\ref{eq:Z_single}), is essential to accurately compute the permutation-cycle distribution $P(l)$, and to perform a meaningful check of our PIMC data.



\section{Results\label{sec:results}}

\begin{figure*}
\centering
\includegraphics[width=0.4\textwidth]{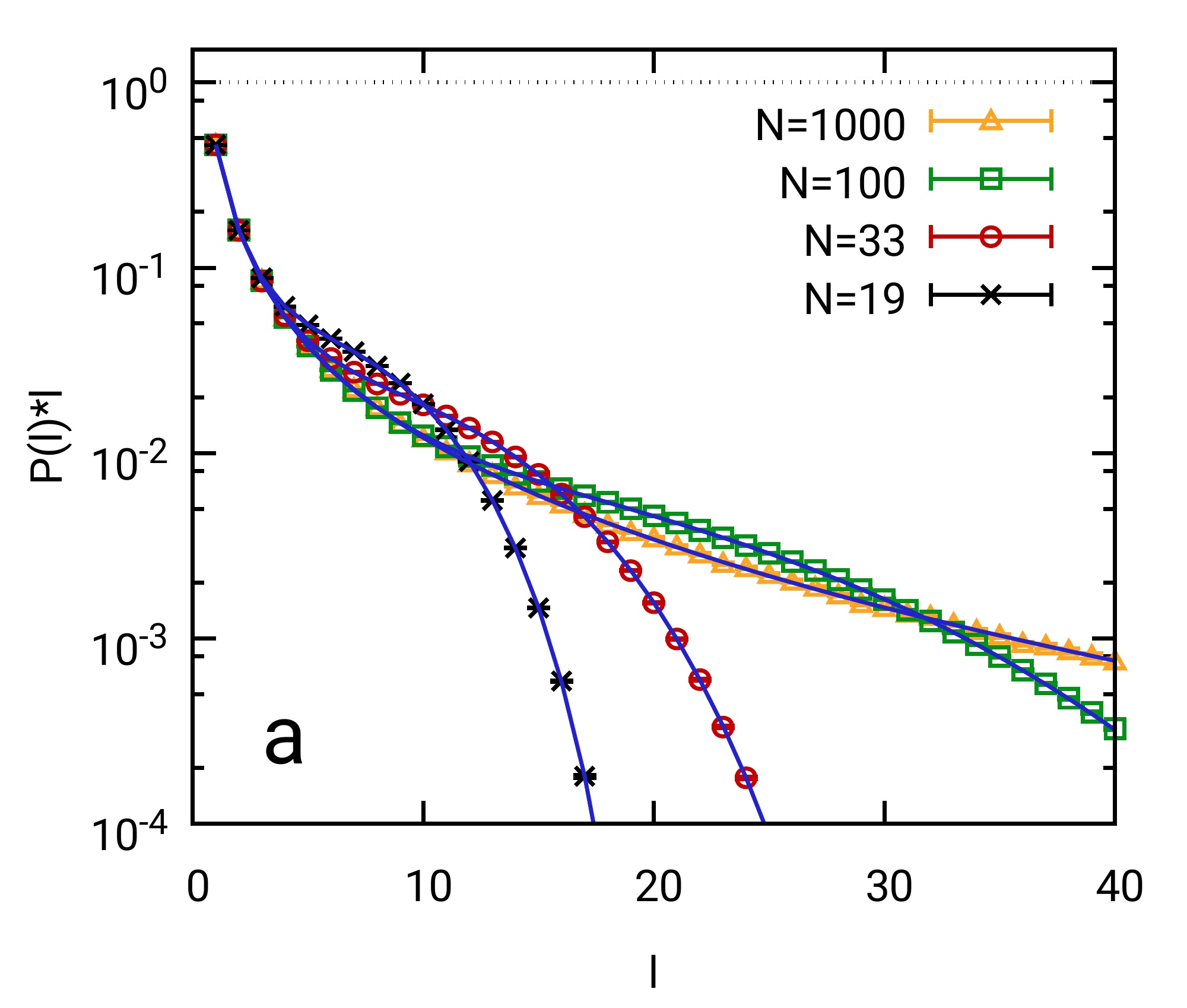}
\includegraphics[width=0.4\textwidth]{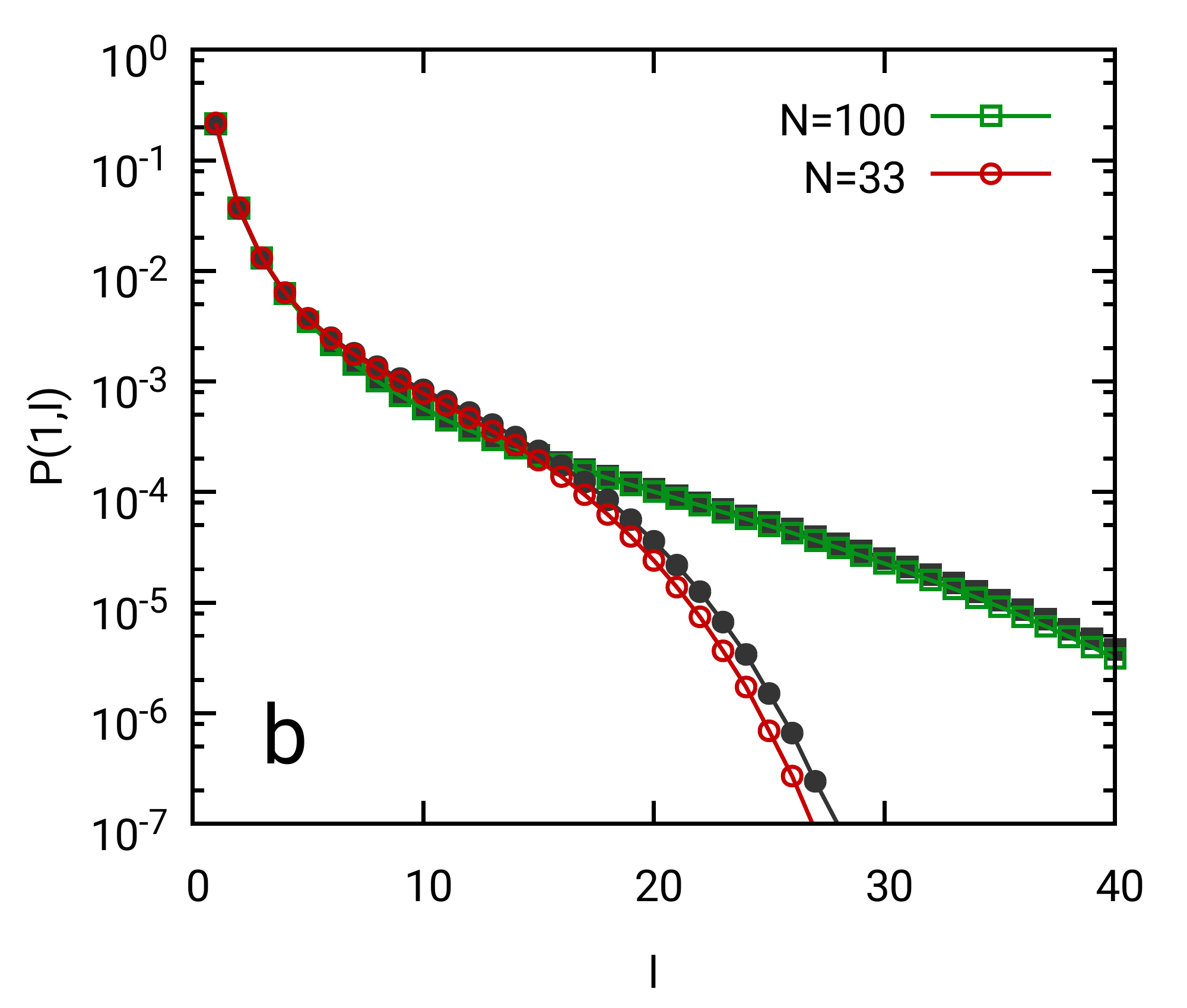}\\
\includegraphics[width=0.4\textwidth]{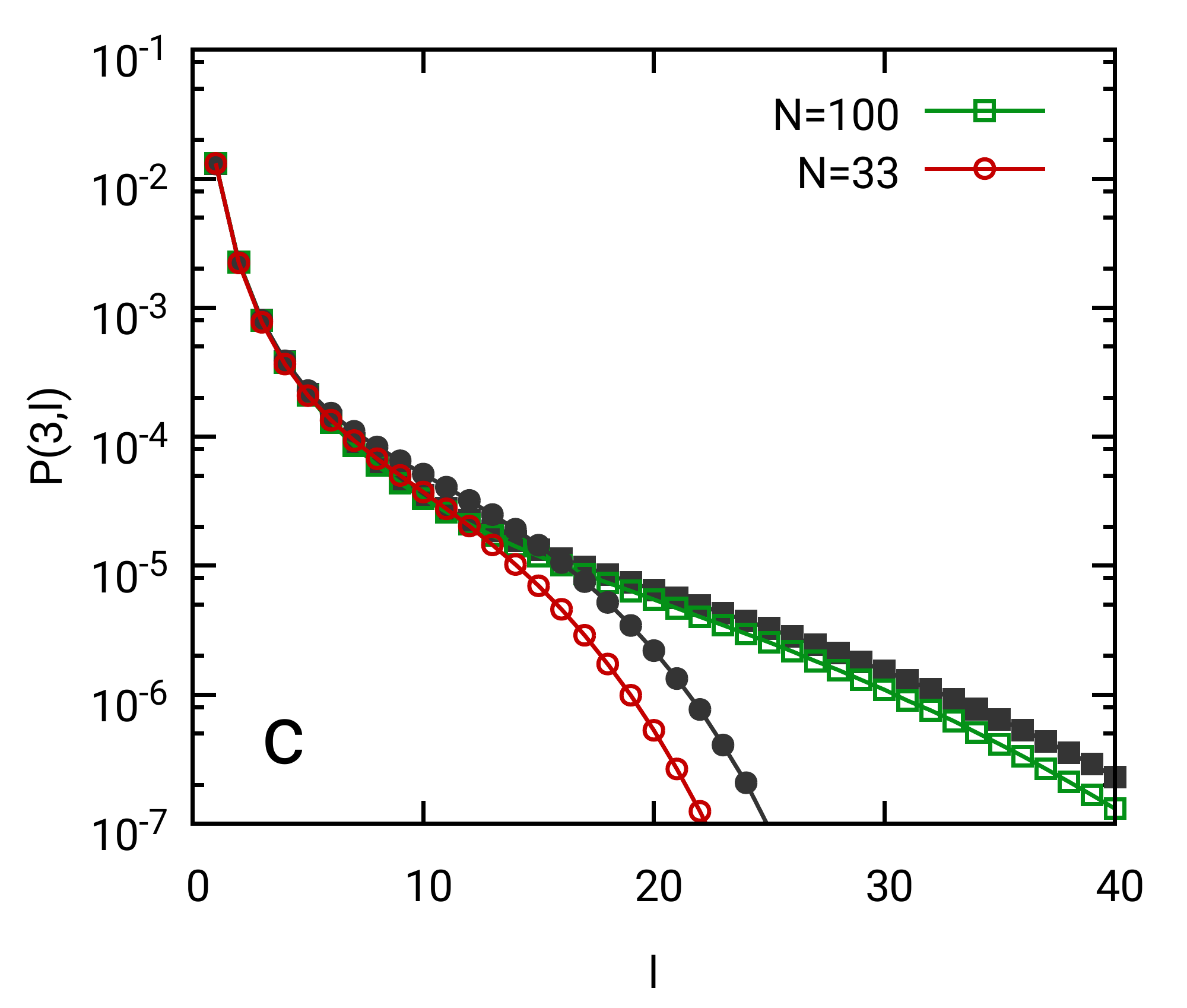}
\includegraphics[width=0.4\textwidth]{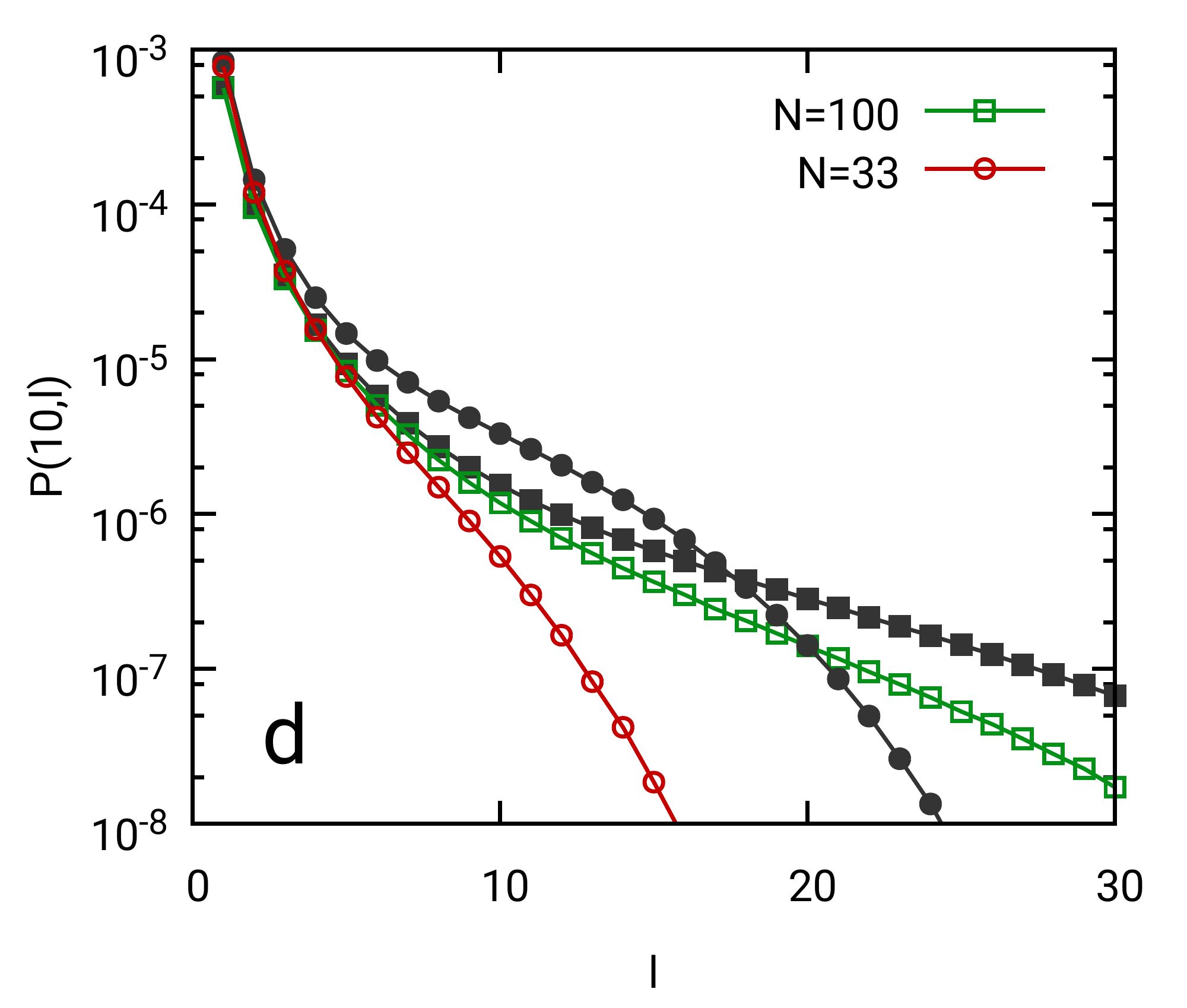}
\caption{\label{fig:column_ideal}
PIMC results for the permutation cycle properties of the ideal Fermi gas at $\theta=0.5$---
Probability of each particle to be involved in a permutation cycle of length $l$, $P(l)l$ (panel a, symbols depict PIMC data and blue lines the analytical result [Eq.~(\ref{eq:analytical_pl})]), and segments of the permutation cycle pair distribution function $P(k,l)$ for $k=1,3,10$ (panels b,c,d), with the colored and grey symbols corresponding to our PIMC results and the uncorrelated analogue $P_\textnormal{u}(k,l)=P(l)P(k)$, respectively.
}
\end{figure*}

\subsection{Ideal Fermi gas\label{sec:results_ideal}}

Let us start our investigation of the permutation-cycle properties by considering $N$ ideal spin-polarized fermions in a three-dimensional periodic box of length $L$.
The corresponding Hamiltonian is simply given by
\begin{eqnarray}
\hat H = -\frac{\hbar^2}{2m} \sum_{k=1}^N \nabla_k^2 \quad ,
\end{eqnarray}
which is readily diagonalised by plane-waves of the form \begin{eqnarray}
\Phi_{\mathbf{k}}(\mathbf{r}) = \langle \mathbf{r}|\mathbf{k}\rangle = \frac{e^{i\mathbf{r}\cdot\mathbf{k}}}{\sqrt{V}} \quad , 
\end{eqnarray}
with $\mathbf{k}=2\pi L^{-1} (a_0,a_1,a_2)^T$ and $a_i\in \mathbb{Z}$,
leading to the partition function in Eq.~(\ref{eq:Z_single}).

In panel a) of Fig.~\ref{fig:column_ideal}, we show the probability of a single particle to be in a permutation-cycle of length $l$, $P(l) l$, for $N=1000$ (yellow triangles), $N=100$ (green squares), $N=33$ (red circles), and $N=19$ (black crosses) ideal fermions at $\theta=0.5$. For completeness, we mention that the density-parameter has been chosen as $r_s=2$ in our PIMC simulations, although the ideal results are independent of $r_s$. The symbols correspond to our PIMC data evaluated as a histogram according to Eq.~(\ref{eq:p}), and the continuous blue lines to the semi-analytical result, Eq.~(\ref{eq:analytical_pl}).
First and foremost, we note the perfect agreement between the PIMC data and the exact result over the entire $l$-range, which is a strong verification of our implementation. An additional comparison for different temperatures can be found in Fig.~\ref{fig:Pl_T}.
Moreover, the $P(l)l$ data for different $N$ are in agreement with each other for small $l$, but there appear large-sized deviations for $l\gtrsim 4$.
These finite-size effects manifest as a steep decrease for large $l$, which is a consequence of the periodicity of the simulation box. More specifically, if we have a permutation-cycle of length $l$ and propose a pair-exchange with a particle that is already involved in the same trajectory, this exchange-cycle is split up into two smaller ones. Therefore, large permutation-cycles are significantly suppressed as compared to the macroscopic system in the thermodynamic limit, $N\to\infty$. Lastly, we mention that $P(l)l$ does not exhibit an exponential decay with $l$ for any depicted system size, even in the case of noninteracting fermions.

\begin{figure}
\centering
\includegraphics[width=0.4\textwidth]{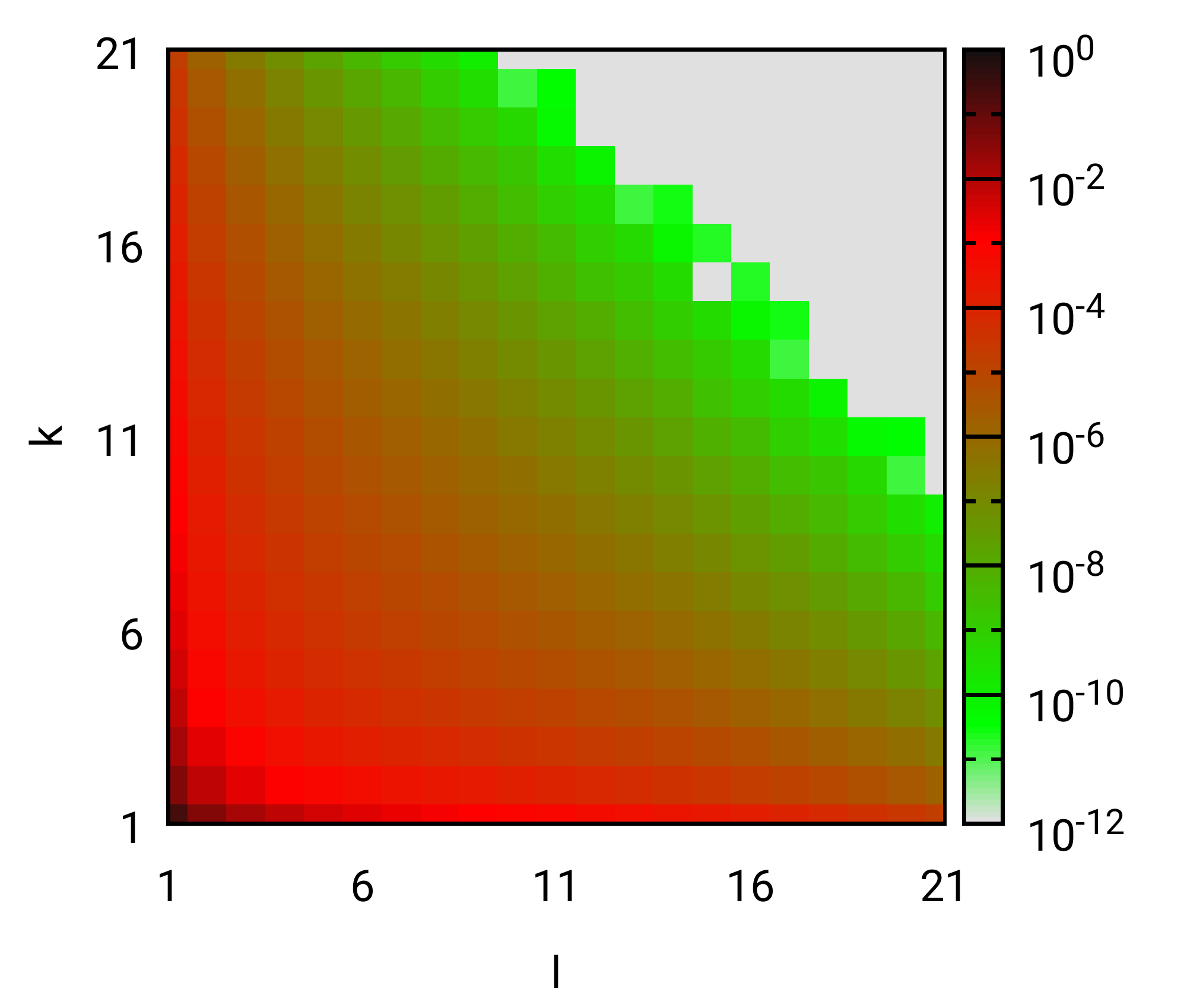}\\
\includegraphics[width=0.4\textwidth]{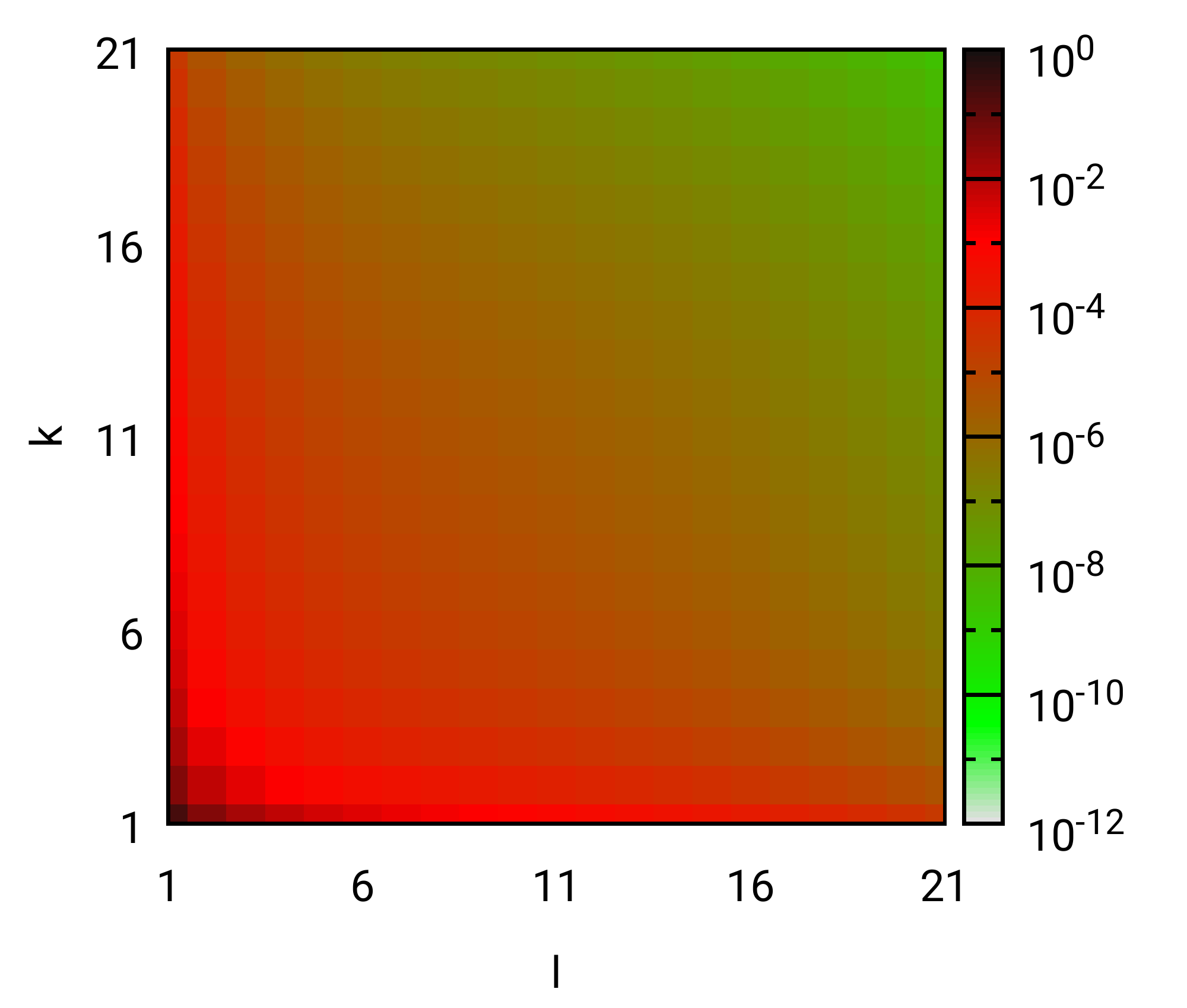}
\caption{\label{fig:pcf_N33_ideal}
Permutation cycle pair distribution function $P(k,l)$ for the ideal Fermi gas with $N=33$ and $\theta=0.5$.
The top and bottom panels correspond to the PIMC results [see Eq.~(\ref{eq:pcf})] and the uncorrelated analogue given by $P_\textnormal{u}(k,l)=P(l)P(k)$, respectively.
}
\end{figure}

In Fig.~\ref{fig:column_ideal} b), we show PIMC data for the permutation-cycle correlation function $P(1,l)$ [see Eq.~(\ref{eq:pcf})] for $N=100$ (green squares) and $N=33$ (red circles), which is a measure for the joint probability to find one single particle and one permutation-cycle of length $l$ within the PIMC simulation at the same time. Again, we observe a perfect agreement of both data sets for small $l$, and significant finite-size effects for $l\gtrsim 5$. The dark grey symbols show the corresponding uncorrelated function $P_\textnormal{u}(1,l)$ [see Eq.~(\ref{eq:pcf_uncorrelated}) in Sec.~\ref{sec:permutation_cycle_theory}], which, for ideal fermions in the thermodynamic limit, should exactly reproduce $P(1,l)$.
While this does indeed (approximately) hold for $N=100$ over the entire depicted $l$-range, there appear sizeable deviations for $N=33$ for $l\gtrsim15$. In fact, it always holds $P(k,l)=0$ for $k+l>N$, which is not the case for the uncorrelated analogue $P_u(k,l)$, and the two quantities deviate even for $l+k<N$, cf.~Fig.~\ref{fig:pcf_N33_ideal}. 
In panels c) and d) of Fig.~\ref{fig:column_ideal}, we show $P(3,l)$ and $P(10,l)$, which exhibit a similar behavior, although the finite-size effects are more pronounced and start for smaller $l$.

For completeness, in Fig.~\ref{fig:pcf_N33_ideal} we show the full $k$- and $l$-dependence of $P(k,l)$ (top) and $P_\textnormal{u}(k,l)$ (bottom) for the case of $N=33$.
In a nutshell, the joint probability to find a permutation-cycle of length $l$ depends on the composition of a configuration $\mathbf{X}$ even for ideal fermions, which is a direct consequence of the finite number of particles.

\subsection{Uniform electron gas\label{sec:ueg_results}}

Let us next consider the case of the uniform electron gas~\cite{dornheim_review,Gill_UEGREVIEW_2011}, which is governed by the Hamiltonian
\begin{eqnarray}
\hat H = -\frac{\hbar^2}{2m} \sum_{k=1}^N \nabla_k^2 
+ \sum_{k=1}^N \sum_{l=k+1}^N W(\mathbf{\hat r}_k,  \mathbf{\hat r}_l) \quad,
\end{eqnarray}
with $W(\mathbf{r},\mathbf{r}')$ being the pair-interaction. To speed up our calculations, we have replaced the usual Ewald summation~\cite{fraser_PRB} with the spherically-averaged potential by Yakub and Ronchi~\cite{Yakub_JCP,Yakub2005}, where the summations in reciprocal and coordinate space can be carried out analytically.
In the moderately coupled warm dense matter regime that is of interest in this work, the effect on the permutation-cycle properties is negligible.

\begin{figure}
\centering
\includegraphics[width=0.4\textwidth]{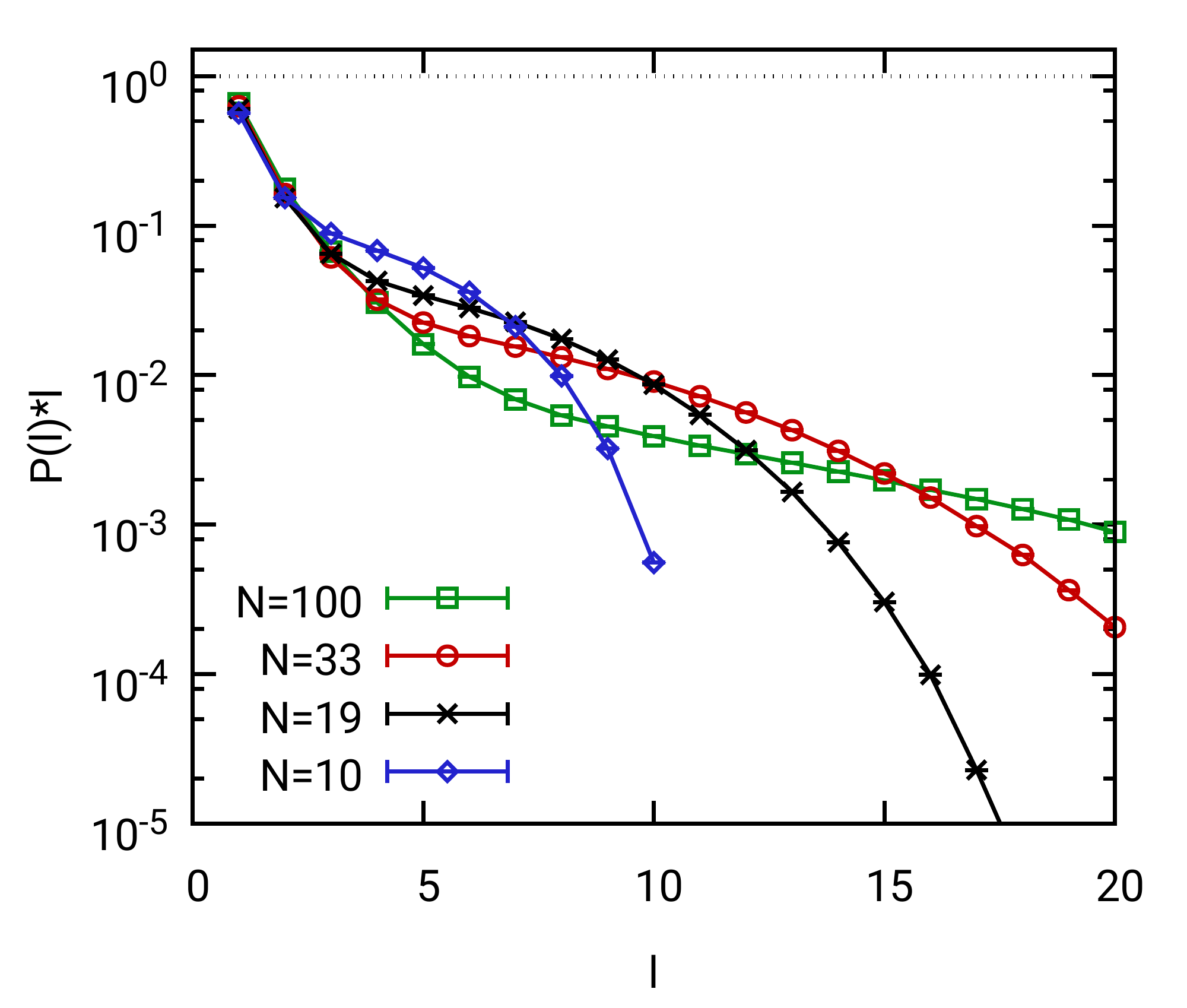}\\
\includegraphics[width=0.4\textwidth]{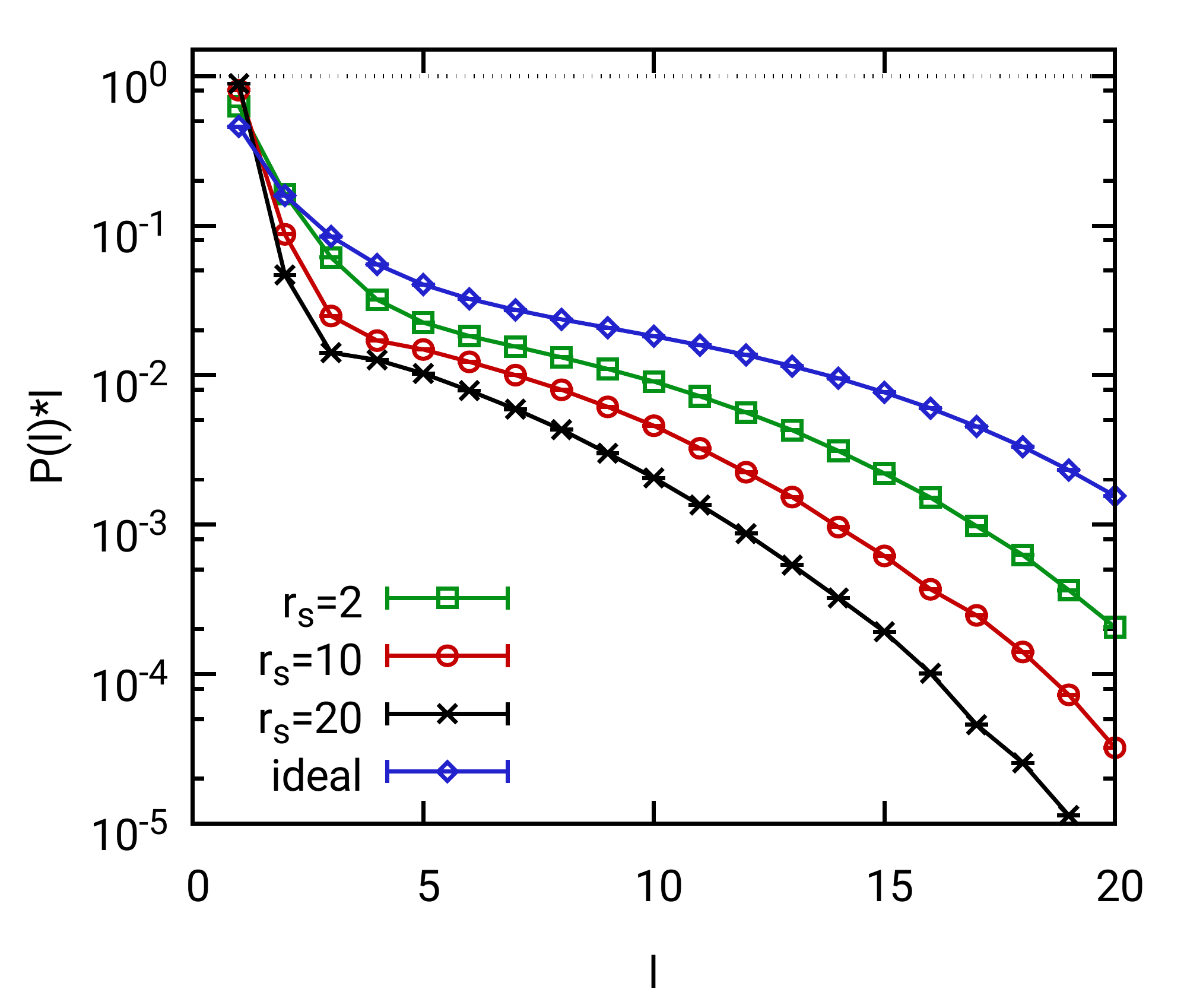}
\caption{\label{fig:Pl_rs_N}
Probability of a particle to be involved in a permutation cycle of length $l$, $P(l)l$,
for the uniform electron gas for different particle numbers $N$ at $r_s=2$ and $\theta=0.5$ (top) and for different density parameters $r_s$ at $N=33$ and $\theta=0.5$ (bottom).
}
\end{figure}

In Fig.~\ref{fig:Pl_rs_N} (top), we show PIMC results for $P(l)l$ for the spin-polarized UEG in the warm dense matter regime at $r_s=2$ and $\theta=0.5$ with $N=100$ (green squares), $N=33$ (red circles), $N=19$ (black crosses), and $N=10$ (blue diamonds). First and foremost, we note that the results are qualitatively very similar to the noninteracing data shown in Fig.~\ref{fig:column_ideal}, including the distinct finite-size effects. Moreover, here, too, we do not find an exponential decay of $P(l)$ with the permutation-cycle length $l$. In the bottom panel, we compare PIMC data for the UEG with $N=33$ electrons at $\theta=0.5$ for $r_s=2$ (green squares), $r_s=10$ (red circles), $r_s=20$ (black crosses), and the noninteracting case (blue diamonds). With increasing $r_s$, the system becomes more sparse and, consequently, the coupling strength increases~\cite{giuliani2005quantum}. The resulting repulsion-induced inter-particle separation leads to a suppression of exchange-effects. Thus, we find an increased probability of single-particle trajectories, whereas permutation-cycles with $l\geq 2$ occur less often as compared to the noninteracting case.

\begin{figure}
\centering
\includegraphics[width=0.4\textwidth]{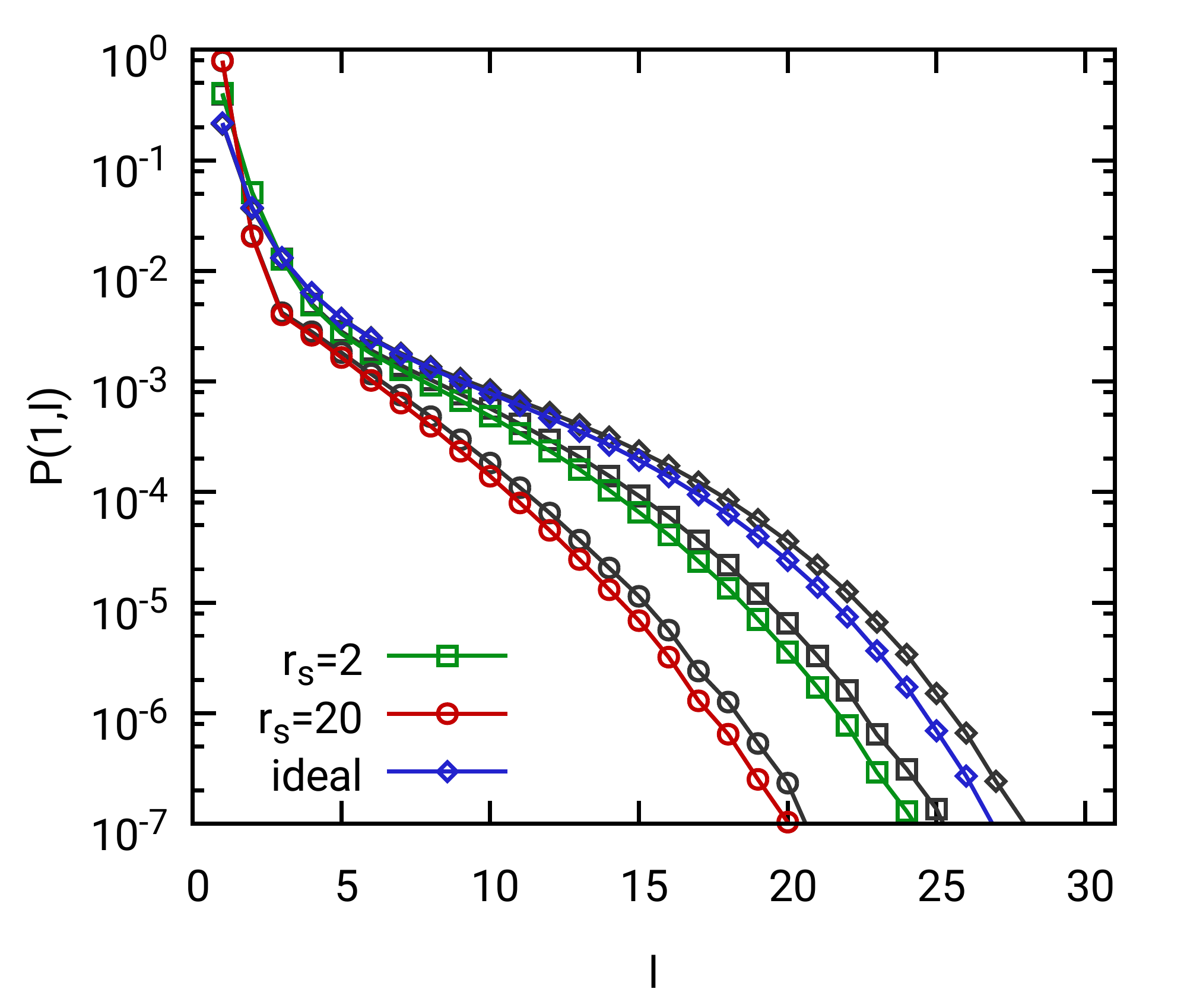}\\
\includegraphics[width=0.4\textwidth]{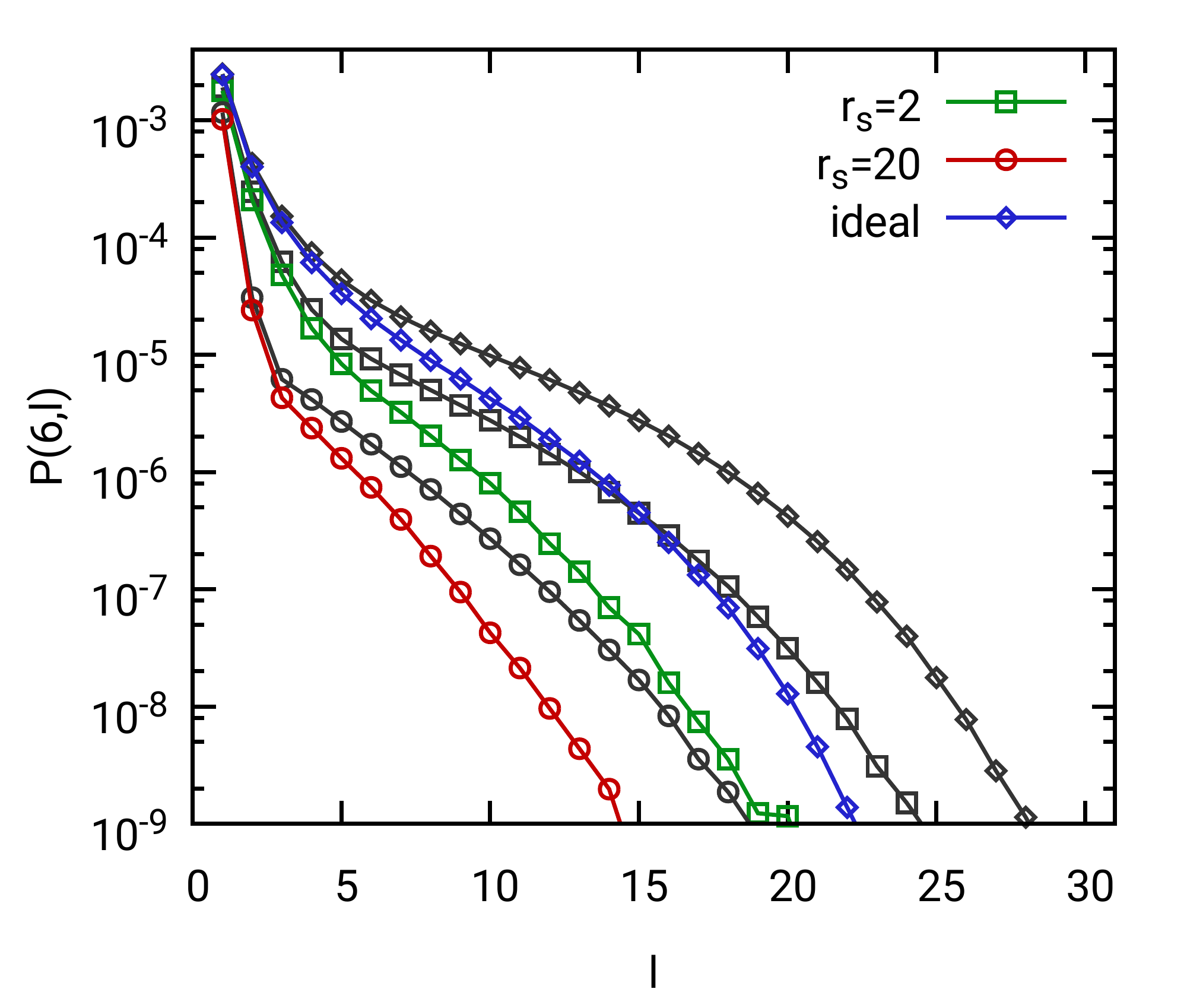}
\caption{\label{fig:column_N33_theta0p5_rs20}
Segments of the permutation cycle pair distribution function $P(k,l)$ for $k=1$ (top) and $k=6$ (bottom) for the uniform electron gas with $N=33$, and $\theta=0.5$ for $r_s=2$ and $r_s=20$. The colored and grey symbols correspond to our PIMC results and the uncorrelated analogue $P_\textnormal{u}(k,l)=P(l)P(k)$, respectively.
}
\end{figure}

Let us next consider the permutation-cycle correlation function $P(k,l)$, which is shown in Fig.~\ref{fig:column_N33_theta0p5_rs20}. In the top panel, we investigate the correlation between single-particles and cycles of length $l$, $P(1,l)$, for $N=33$ and $\theta=0.5$ for $r_s=2$ (green squares), $r_s=20$ (red circles), and the ideal case (blue diamonds). We note that all three curves exhibit a very similar behaviour. The dark grey symbols correspond to the uncorrelated joint probability $P_\textnormal{u}(k,l)$, cf.~Eq.~(\ref{eq:pcf_uncorrelated}). Remarkably, we find that independent of the coupling strength, $P_\textnormal{u}(k,l)$ and $P(k,l)$ are approximately identical for small $l$, which means that the probabilities to find such permutation-cycles are not correlated. In addition, the deviation between the grey and coloured symbols for larger $l$ is qualitatively identical for all three curves. This is a strong indication that the observed correlations in the exchange-cycles are a direct result of the finite number of electrons, and not due to the Coulomb repulsion, even for the strongly coupled case ($r_s=20$).
In the bottom panel, we show the same information for $P(6,l)$.
While the deviations between $P$ and $P_\textnormal{u}$ start to appear even at $l=2$, this difference is again already qualitatively present for the noninteracting case, which means that here, too, Coulomb correlations due not constitute the dominating effect.

\begin{figure*}
\centering
\includegraphics[width=0.4\textwidth]{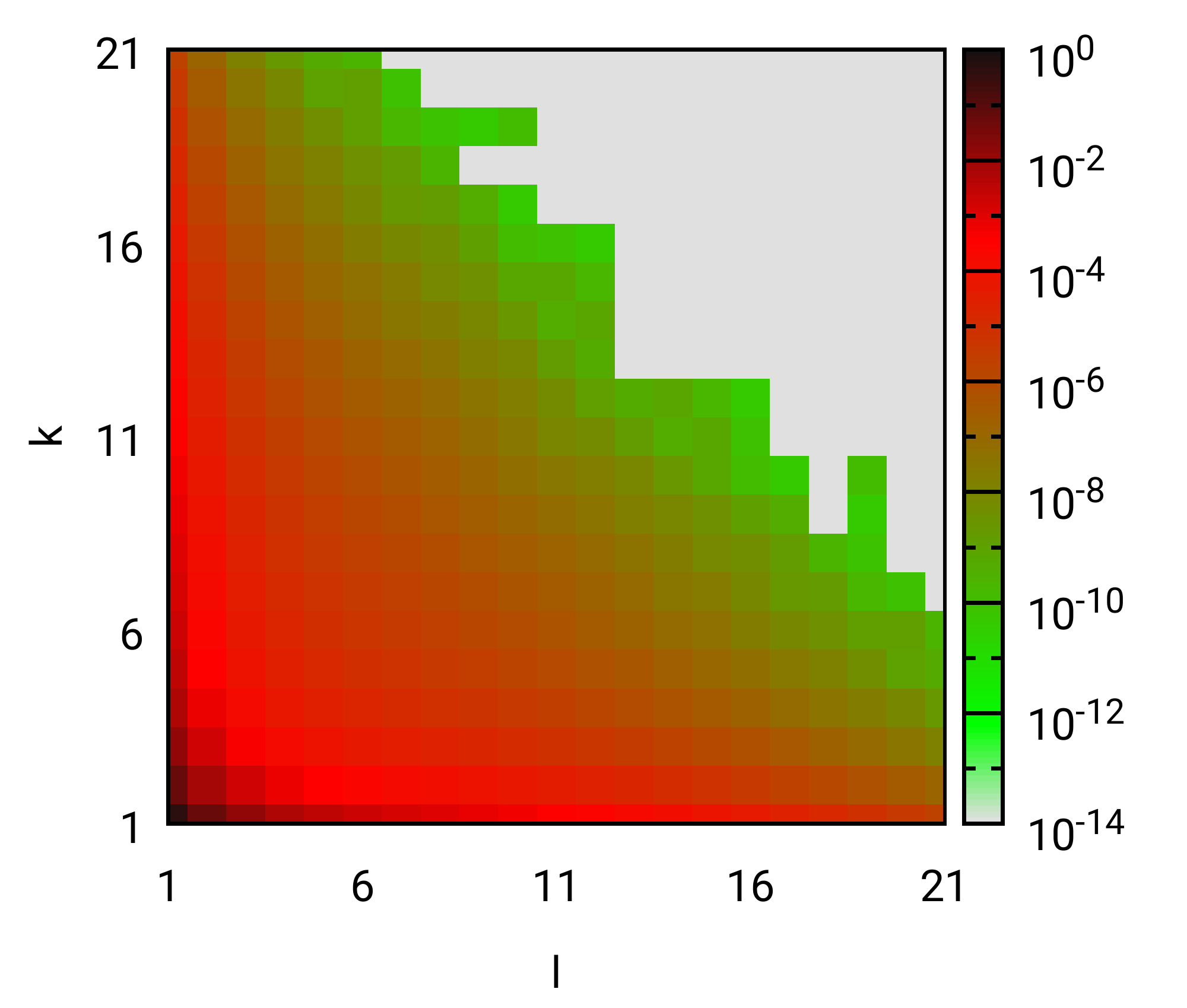}
\includegraphics[width=0.4\textwidth]{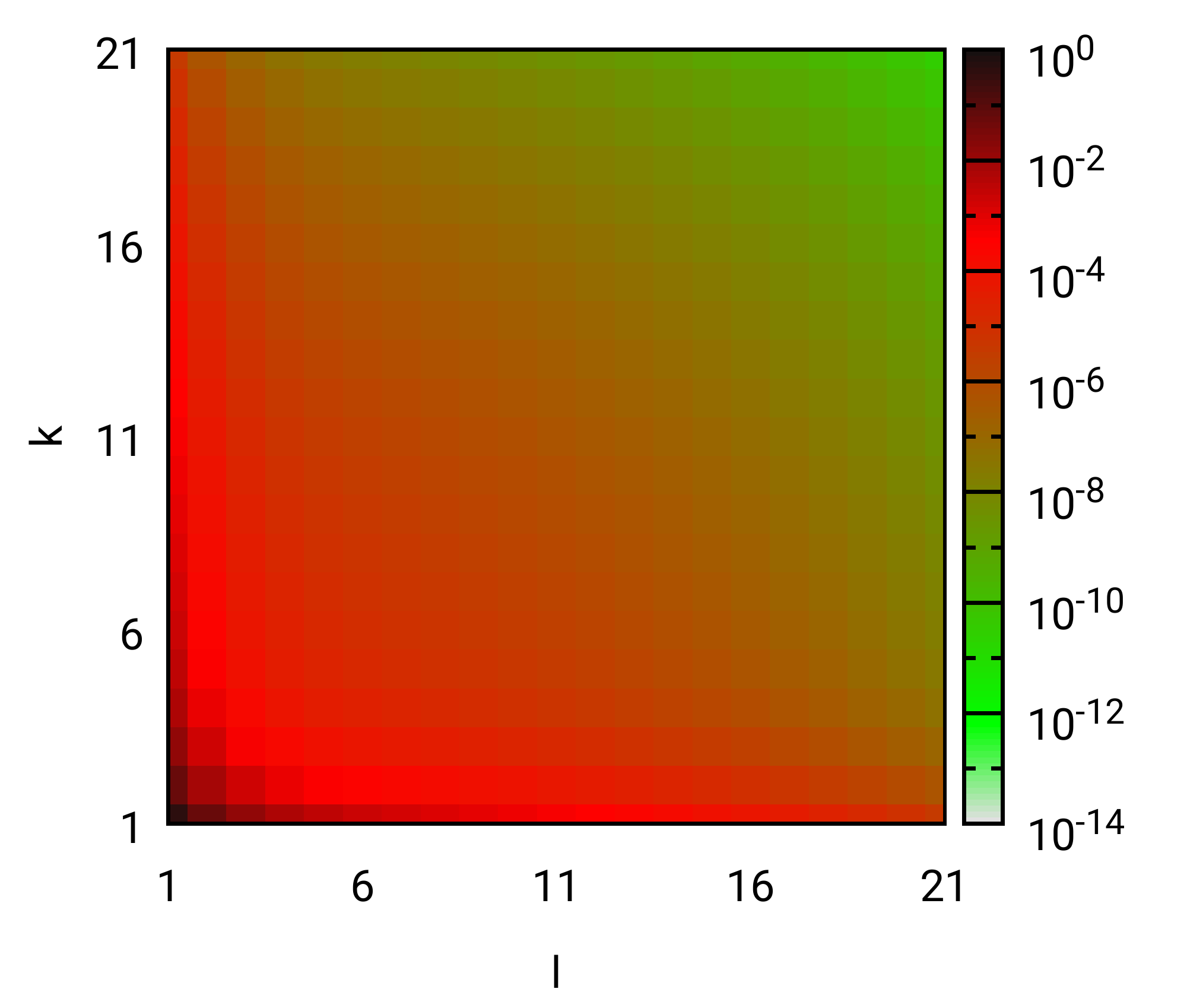}\\
\includegraphics[width=0.4\textwidth]{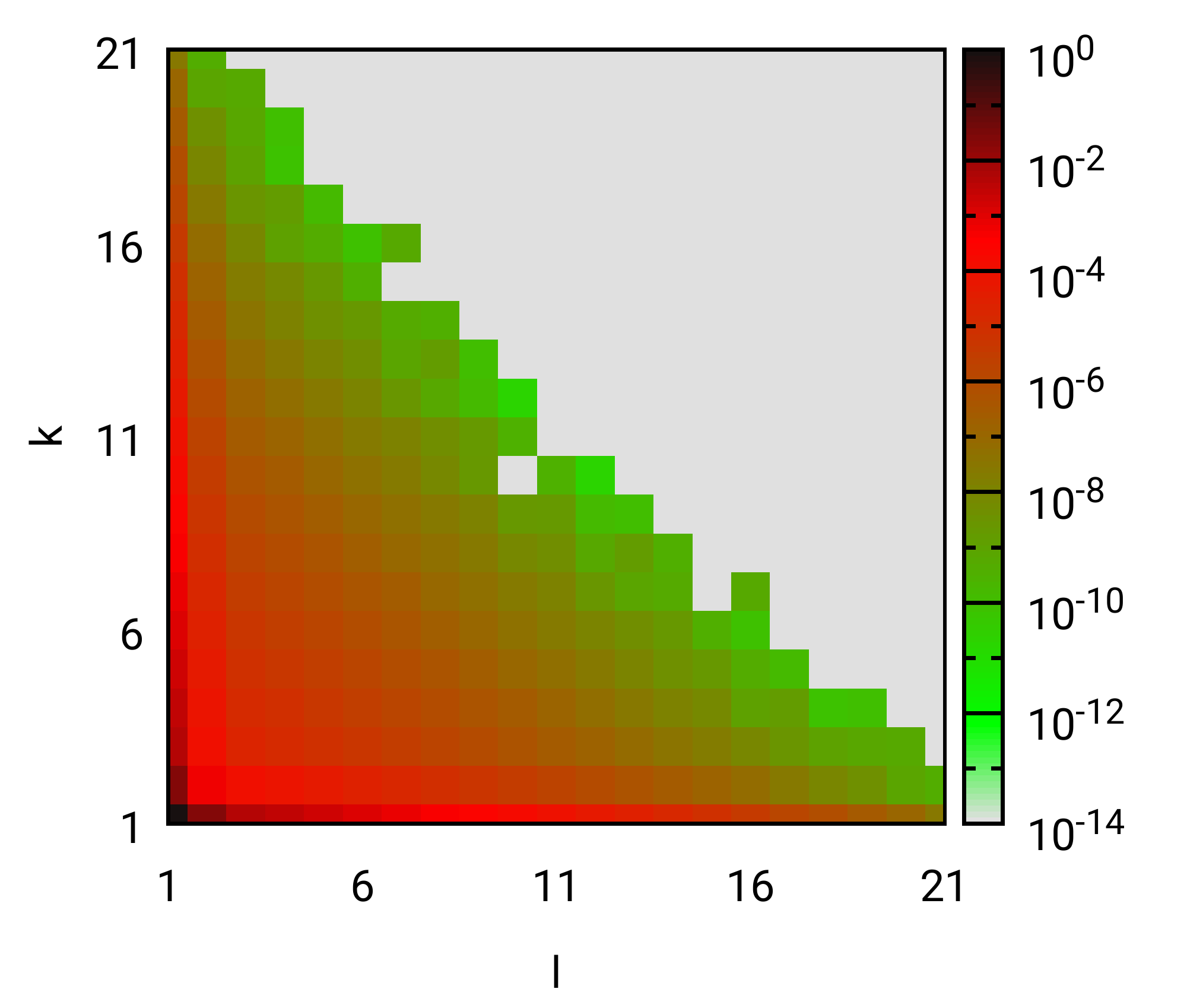}
\includegraphics[width=0.4\textwidth]{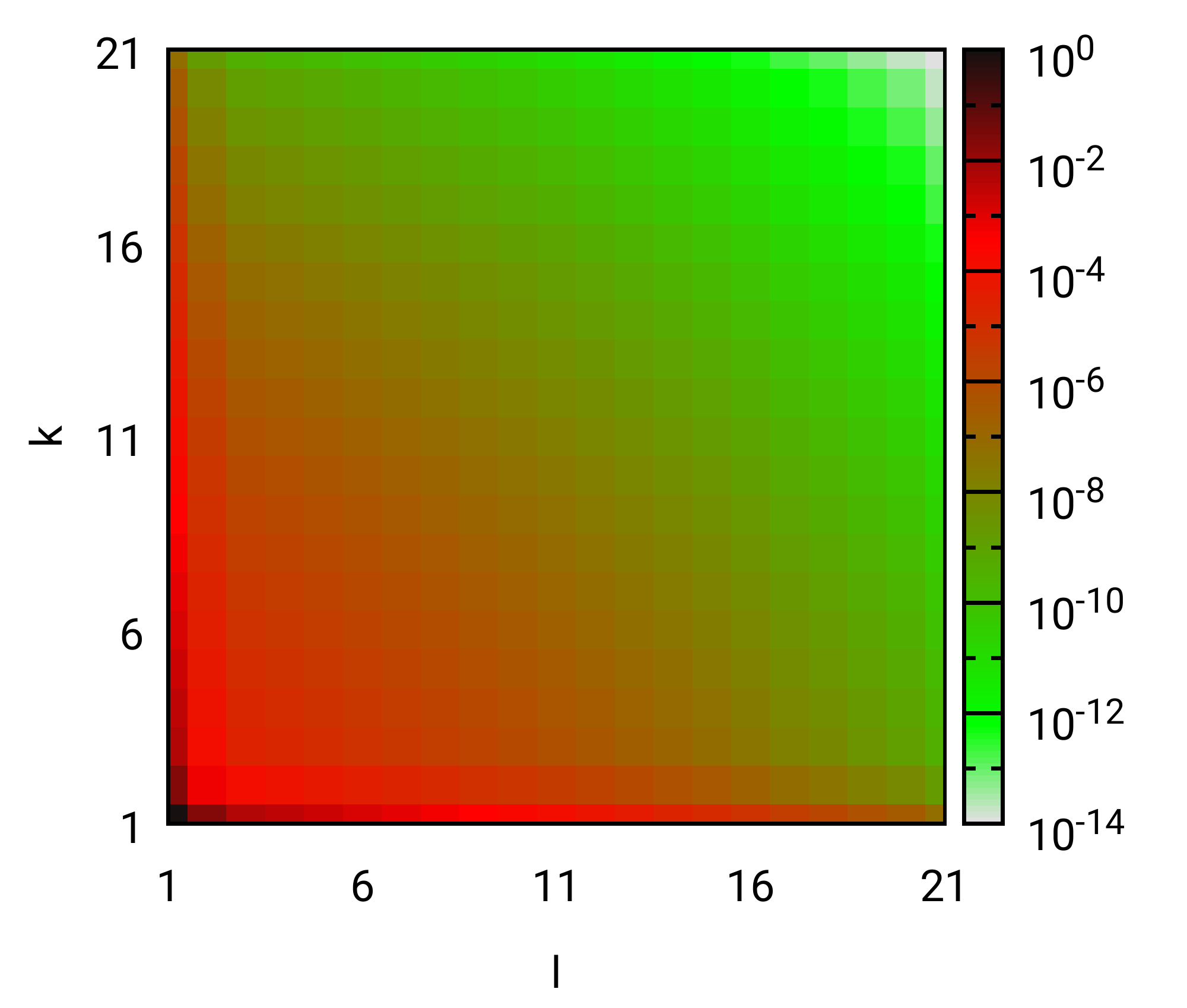}
\caption{\label{fig:pcf_N33_theta0p5_rs20}
Permutation cycle pair distribution function $P(k,l)$ for the uniform electron gas with $N=33$, and $\theta=0.5$ for $r_s=2$ (top row) and $r_s=20$ (bottom row).
The left and right panels correspond to the PIMC results [see Eq.~(\ref{eq:pcf})] and the uncorrelated analogue given by $P_\textnormal{u}(k,l)=P(l)P(k)$, respectively.
}
\end{figure*}

Finally, in Fig.~\ref{fig:pcf_N33_theta0p5_rs20}, we show the full $l$- and $k$-dependence of $P(l,k)$ (left column) and $P_\textnormal{u}(l,k)$ (right column) for $r_s=2$ (top row) and $r_s=20$ (bottom row). Firstly, the grey areas in the top-right corner of the $P$-plots directly follow from the fact that $P(l,k)=0$ for $l+k>N$, see the corresponding discussion in Sec.~\ref{sec:results_ideal}. Apart from this, the uncorrelated function relatively accurately reproduces the real joint probability from our PIMC simulations for small $k$ and $l$, whereas the deviations become more manifest for larger permutation-lengths due to the finite system size.

\begin{figure*}
\centering
\includegraphics[width=0.4\textwidth]{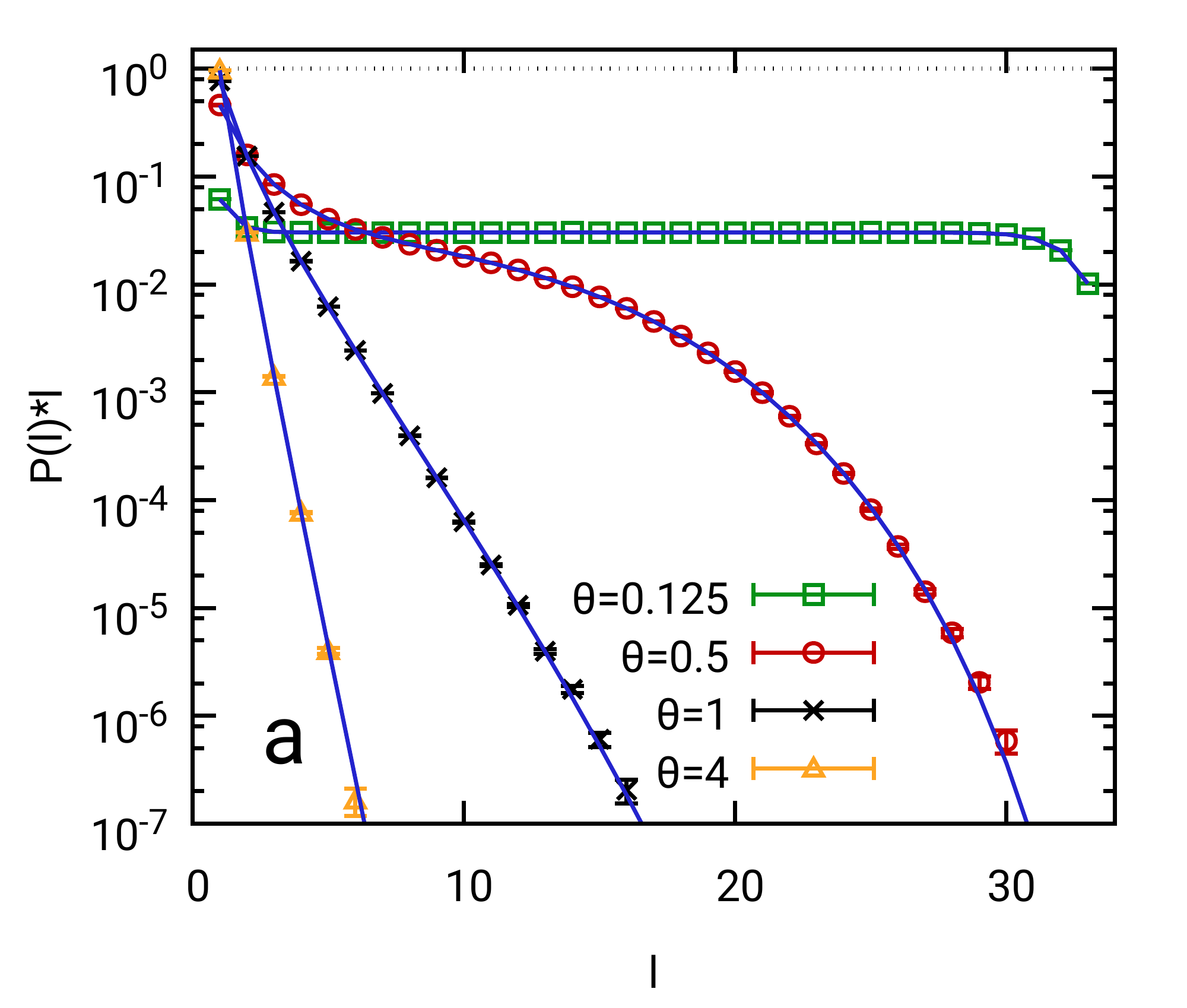}
\includegraphics[width=0.4\textwidth]{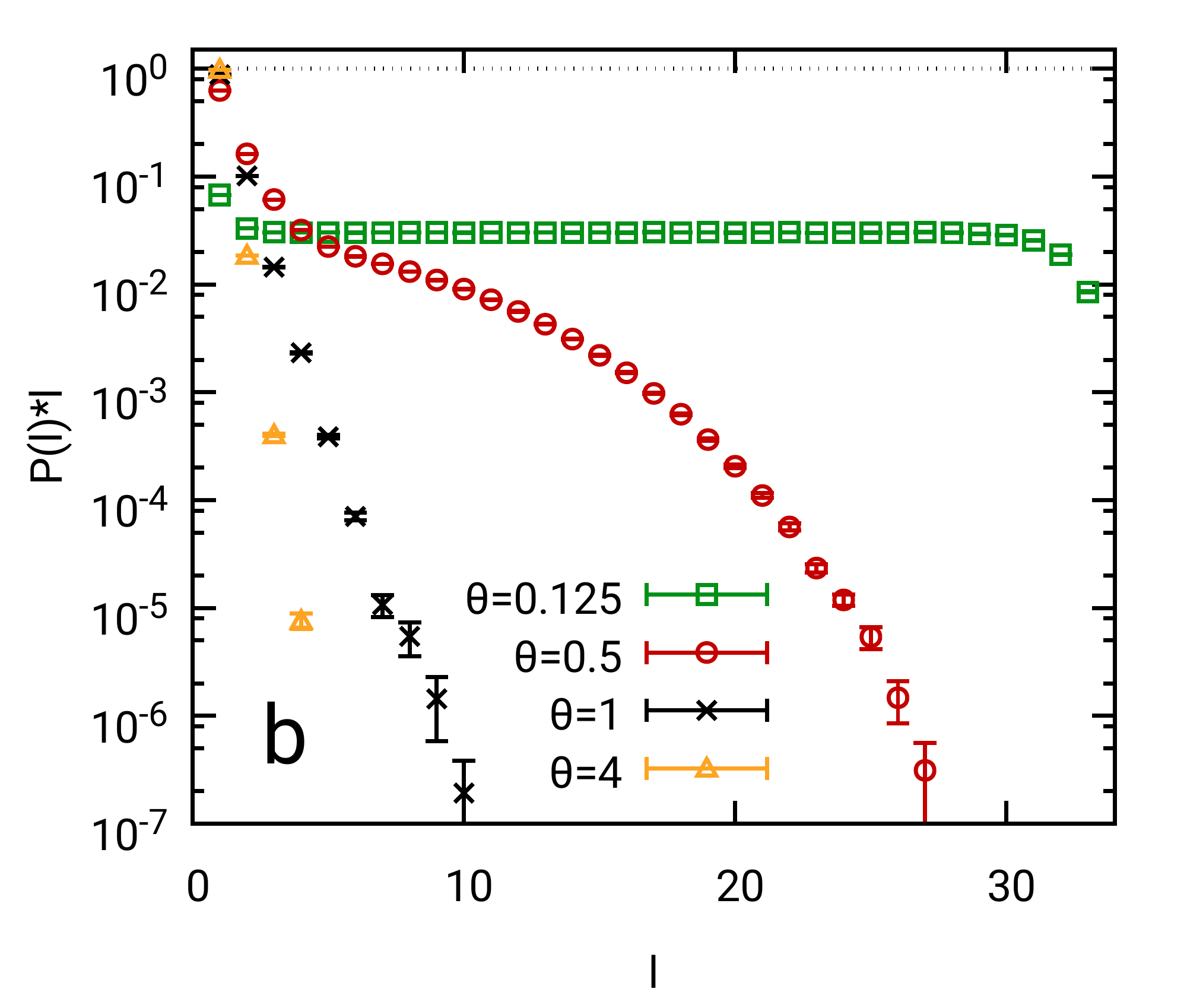}\\
\includegraphics[width=0.4\textwidth]{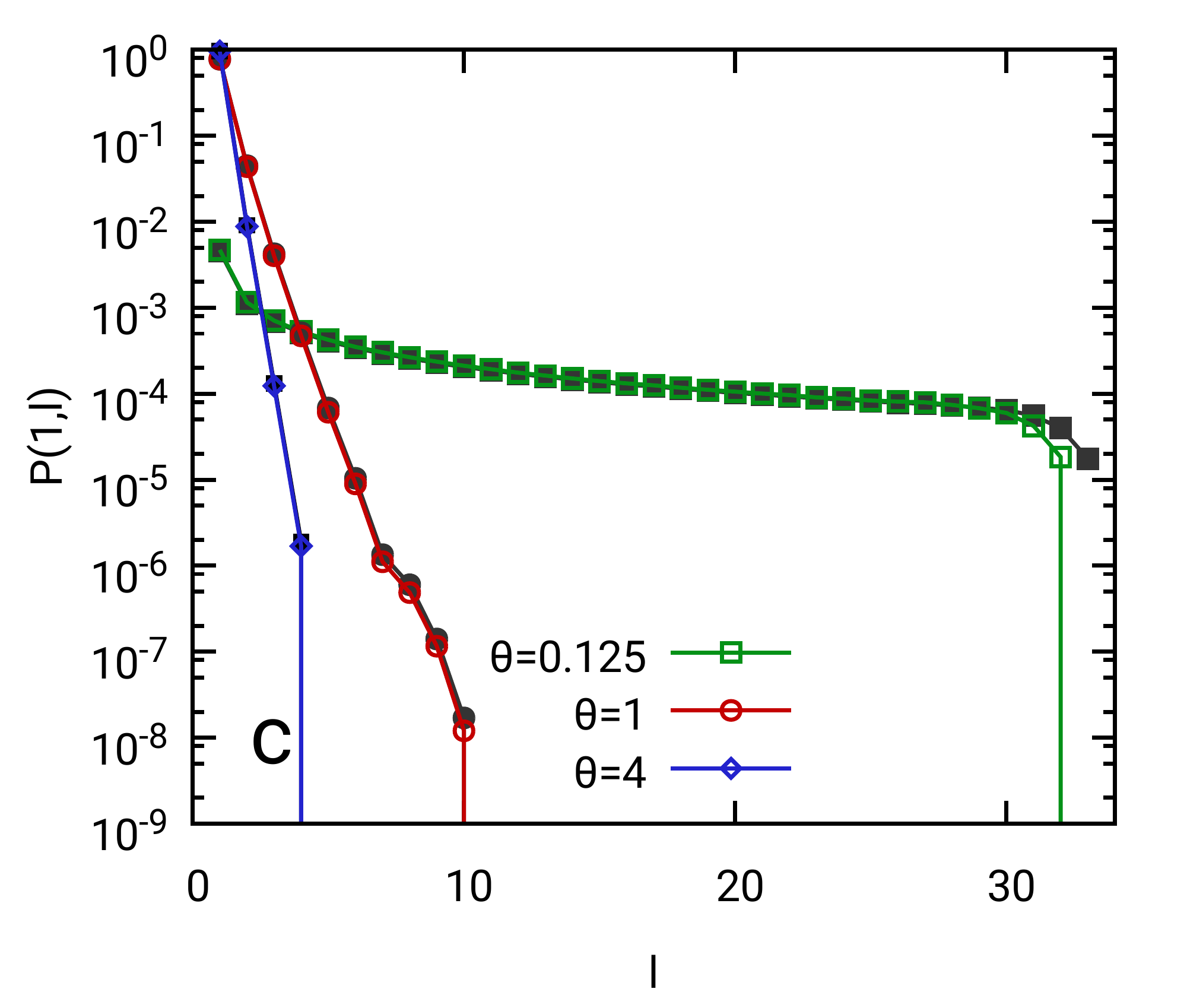}
\includegraphics[width=0.4\textwidth]{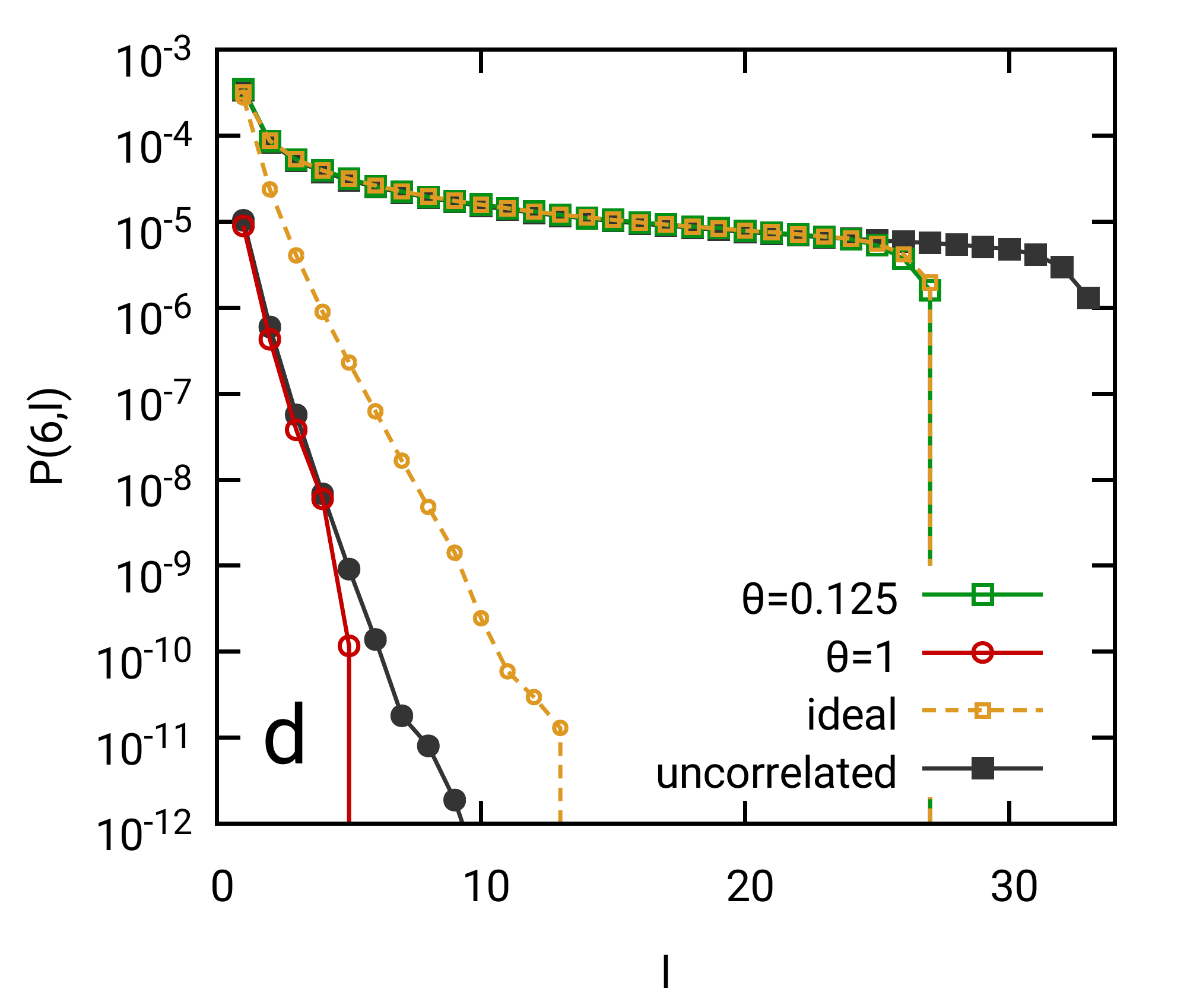}
\caption{\label{fig:Pl_T}
PIMC results for the permutation cycle properties of the ideal Fermi gas and the UEG at $r_s=2$ with $N=33$---Probability of each particle to be involved in a permutation cycle of length $l$, $P(l)l$ for the ideal system (panel a, symbols depict PIMC data and blue lines the analytical result [Eq.~(\ref{eq:analytical_pl})]), 
and the UEG (panel b).
Segments of the permutation cycle pair distribution function $P(k,l)$ of the UEG for $k=1,6$ (panels c and d), with the colored, yellow and grey symbols corresponding to our PIMC results for the UEG and ideal system, and the uncorrelated analogue $P_\textnormal{u}(k,l)=P(l)P(k)$ for the UEG, respectively.
}
\end{figure*}

Let us conclude our analysis of the permutation-cycle properties of PIMC simulations of the warm dense electron gas with an investigation of the temperature-dependence, which is shown in Fig.~\ref{fig:Pl_T}.

In panel a), we show PIMC results for $P(l)l$ for a system of $N=33$ ideal fermions at $\theta=0.125$ (green squares), $\theta=0.5$ (red circles), $\theta=1$ (black crosses), and $\theta=4$ (yellow triangles). Again, the continuous blue lines correspond to the semi-analytical solution [see Eq.~(\ref{eq:analytical_pl})] and are in perfect agreement with our PIMC data over the entire $l$-range for all four temperatures. At $\theta=0.125$, the probability of a particle to be included in a permutation-cycle of length $l$ is almost independent of $l$, resulting in a nearly flat curve. In fact, in the ground state limit, $\theta\to 0$, this probability becomes exactly constant, $P(l)l = 1/N$, which leads to a vanishing average sign $S$ in a PIMC simulation~\cite{krauth_book_2006statistical}. Therefore, the associated statistical uncertainty, Eq.~(\ref{eq:fsp_error}), diverges and PIMC simulations are even theoretically impossible.
With increasing temperature, $P(l)l$ exhibits an increasingly steep descent, whereas the probability to find single-particle trajectories is larger. This is a direct consequence of the smaller extension of the single-particle wave function, cf.~the discussion of Fig.~\ref{fig:snapshots} in Sec.~\ref{sec:permutation_cycle_theory}.
Moreover, we note that, at $\theta=1$ and $\theta=4$, $P(l)l$ does exhibit an approximately exponential decay.

In panel b) we show the same information, but for an interacting electron gas at $r_s=2$, i.e., at a metallic density in the warm dense matter regime. First and foremost, we stress the qualitative similarity to the noninteracting case shown in panel a), in particular for the lowest temperature. In contrast, for $\theta=1$ and $\theta=4$, $P(l)$ does exhibit an exponential decay similar to the ideal case, but with a significantly steeper slope. Therefore, somewhat counter-intuitively, Coulomb-correlation effects are more visible in the permutation-cycle probability at larger temperature, where the system is more weakly coupled.

Finally, in the bottom row of Fig.~\ref{fig:Pl_T} we compare the permutation-cycle correlation function $P(k,l)$ for different temperatures.
In panel c), we show PIMC results for $P(1,l)$ of the UEG at $\theta=0.125$ (green squares) and $\theta=1$ (red circles).
In addition, the corresponding dark grey and yellow symbols depict results for $P_\textnormal{u}(1,l)$ and PIMC results for $P(1,l)$ for the noninteracting system at the same conditions, respectively.
At the lower temperature, all three curves approximately coincide for $l\lesssim 30$, i.e., almost over the entire $l$-range. Only for large permutation-lengths, finite-size effects lead to a deviation towards the uncorrelated joint probability $P_\textnormal{u}(1,l)$, whereas the data for the UEG and the noninteracing system are still in agreement. Evidently, the latter fact is a direct consequence of $P(l)l$ being almost flat in both cases.
In contrast, at $\theta=1$, $P(1,l)$ and $P_\textnormal{u}(1,l)$ are still in good agreement, while the ideal results significantly deviate for all $l$.

Lastly, panel d) shows results for $P(6,l)$ and $P_\textnormal{u}(6,l)$ for the same conditions, with similar results albeit distinctly larger finite-size effects, as it is expected.

\subsubsection{Relative importance of permutation cycle correlations}
\begin{figure}
\centering
\includegraphics[width=0.4\textwidth]{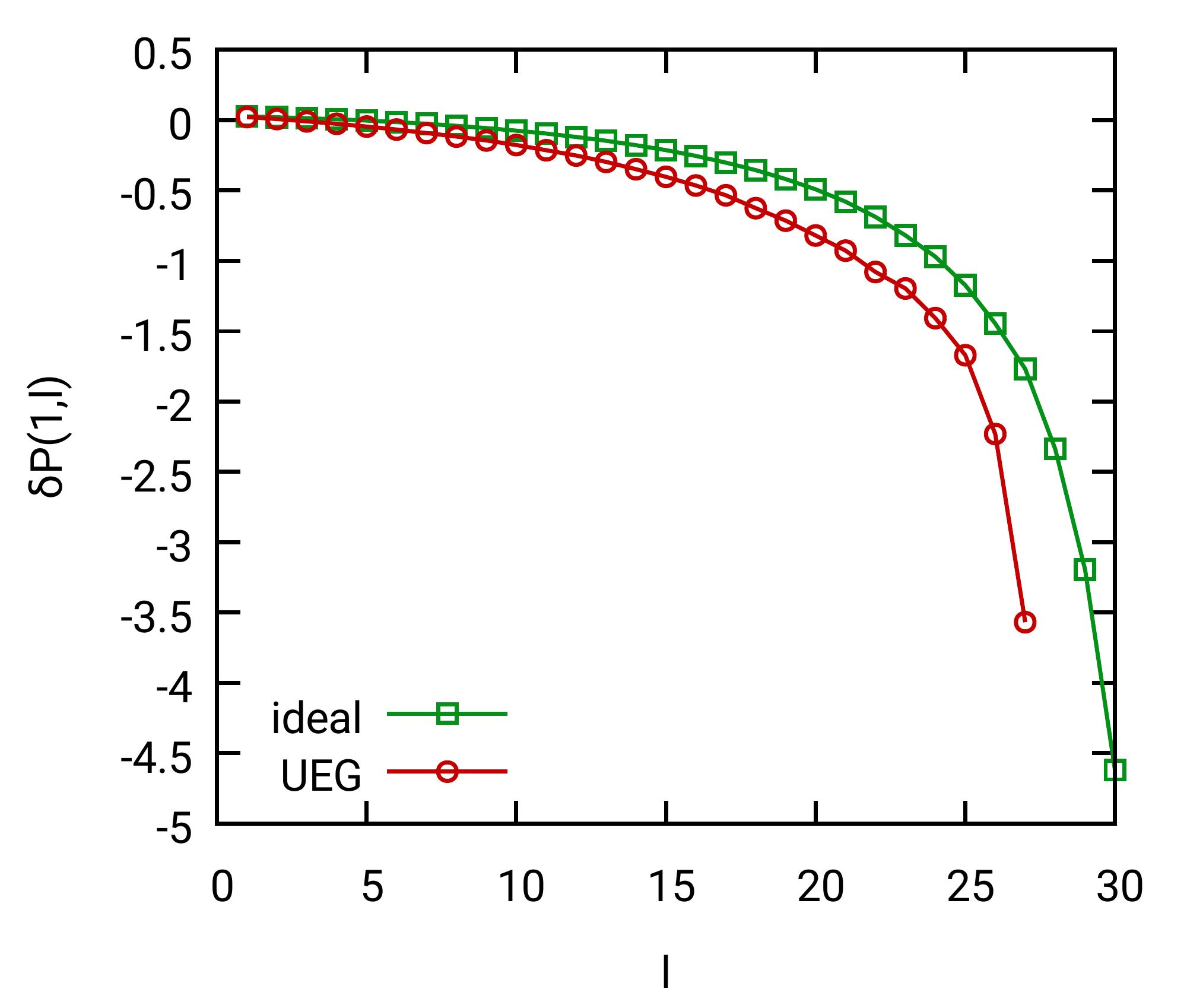}\\
\includegraphics[width=0.4\textwidth]{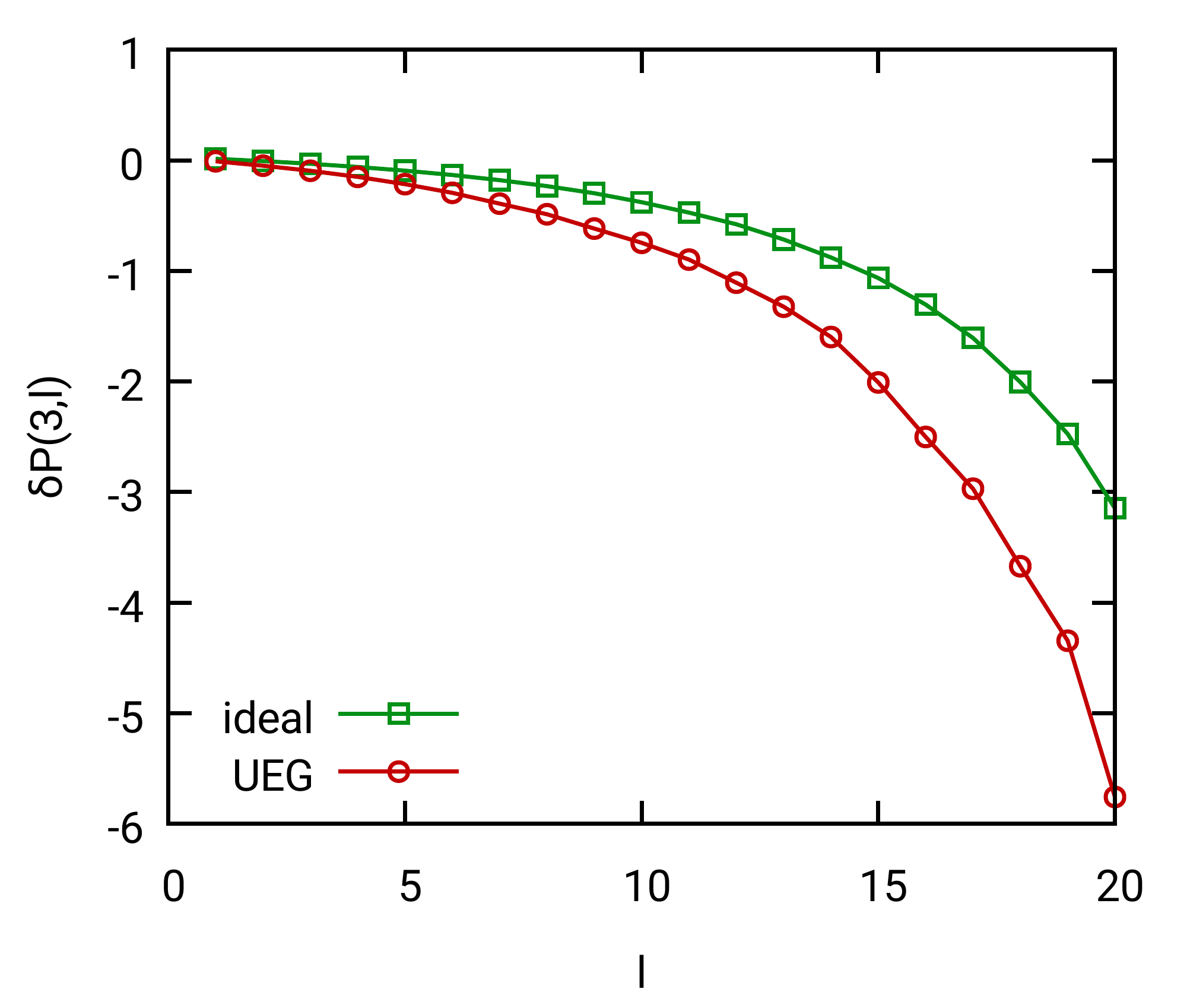}
\caption{\label{fig:deltaa}
Relative deviation $\delta P(l,k)$ [see Eq.~(\ref{eq:deltaa})] between $P(l,k)$ and $P_\textnormal{u}(l,k)=P(l)P(k)$ for $N=33$ particles at $r_s=2$ and $\theta=0.5$. The green squares and red circles correspond to ideal fermions and the uniform electron gas, respectively.
}
\end{figure}
To further quantify the importance of correlations between permutation cycles, we consider the relative deviation between $P(l,k)$ and $P_\textnormal{u}(l,k)$,
\begin{eqnarray}\label{eq:deltaa}
\delta P(l,k) = \frac{P(l,k)-P(l)P(k)}{P(l,k)} \quad .
\end{eqnarray}
The results for Eq.~(\ref{eq:deltaa}) are shown in Fig.~\ref{fig:deltaa} where we plot $\delta P(1,l)$ (top) and $\delta P(3,l)$ (bottom) for $N=33$ particles with $r_s=2$ and $\theta=0.5$. The green squares correspond to the case of ideal fermions and the red circles to the uniform electron gas. In both plots, we observe a remarkably small deviation for small permutation cycles, as it is expected. In contrast, for large cycles ($l\sim20$) there appear significant relative deviations of several hundred percent. This illustrates that the previously observed qualitative agreement between $P(l,k)$ and $P_\textnormal{u}(l,k)$ does not mean that correlation effects of the permutation properties within a PIMC simulation are not important.

\subsection{$2D$ quantum dot\label{sec:itsatrap}}

Let us conclude this paper by extending our considerations to an inhomogeneous system. For this purpose, we simulate electrons in a two-dimensional harmonic trap, which are governed by the Hamiltonian
\begin{eqnarray}\label{eq:trap}
\hat H = -\frac{1}{2} \sum_{k=1}^N \nabla^2_k + \frac{1}{2} \sum_{k=1}^N \mathbf{r}_k^2 + \sum_{k=1}^N \sum_{l=k+1}^N \frac{\lambda}{|\mathbf{r}_l - \mathbf{r}_k|} \quad ,
\end{eqnarray}
where we have used oscillator units (i.e., distances are given in units of $l_0=\sqrt{\hbar/(m\omega)}$ and energies in units of $E_0=\hbar\omega$), and $\lambda$ denotes the dimensionless coupling constant.
Such a system is often used as a simple model for electrons in a quantum dot~\cite{reimann_RevModPhys2002,dornheim_CPP2016,kylanpaa_PhysRevB2017}, and constitutes a useful benchmark for the development of quantum Monte Carlo methods~\cite{egger_PRL1998,schoof_CPP2011,dornheim_NJP2015}.

\begin{figure*}
\centering
\includegraphics[width=0.4\textwidth]{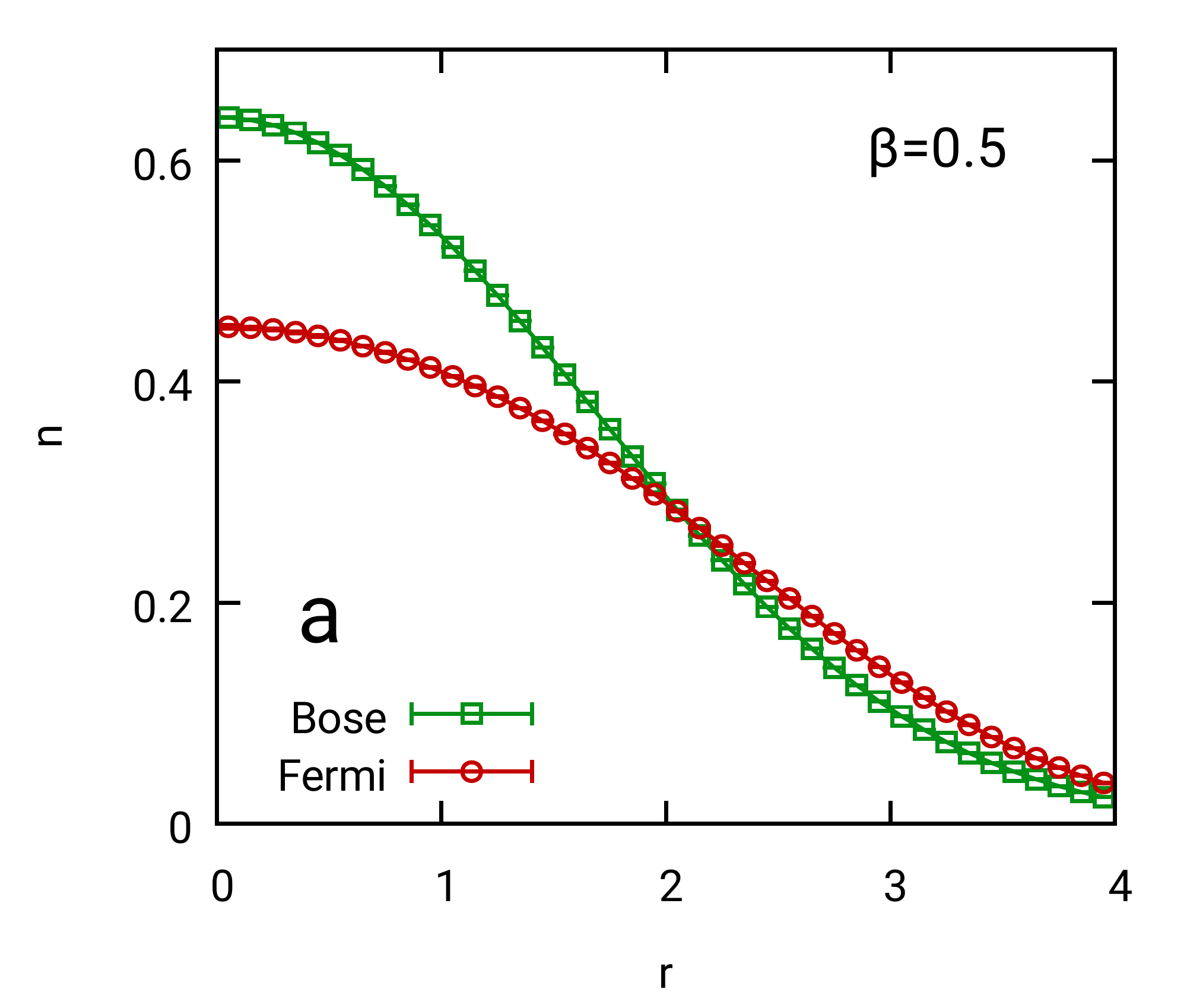}
\includegraphics[width=0.4\textwidth]{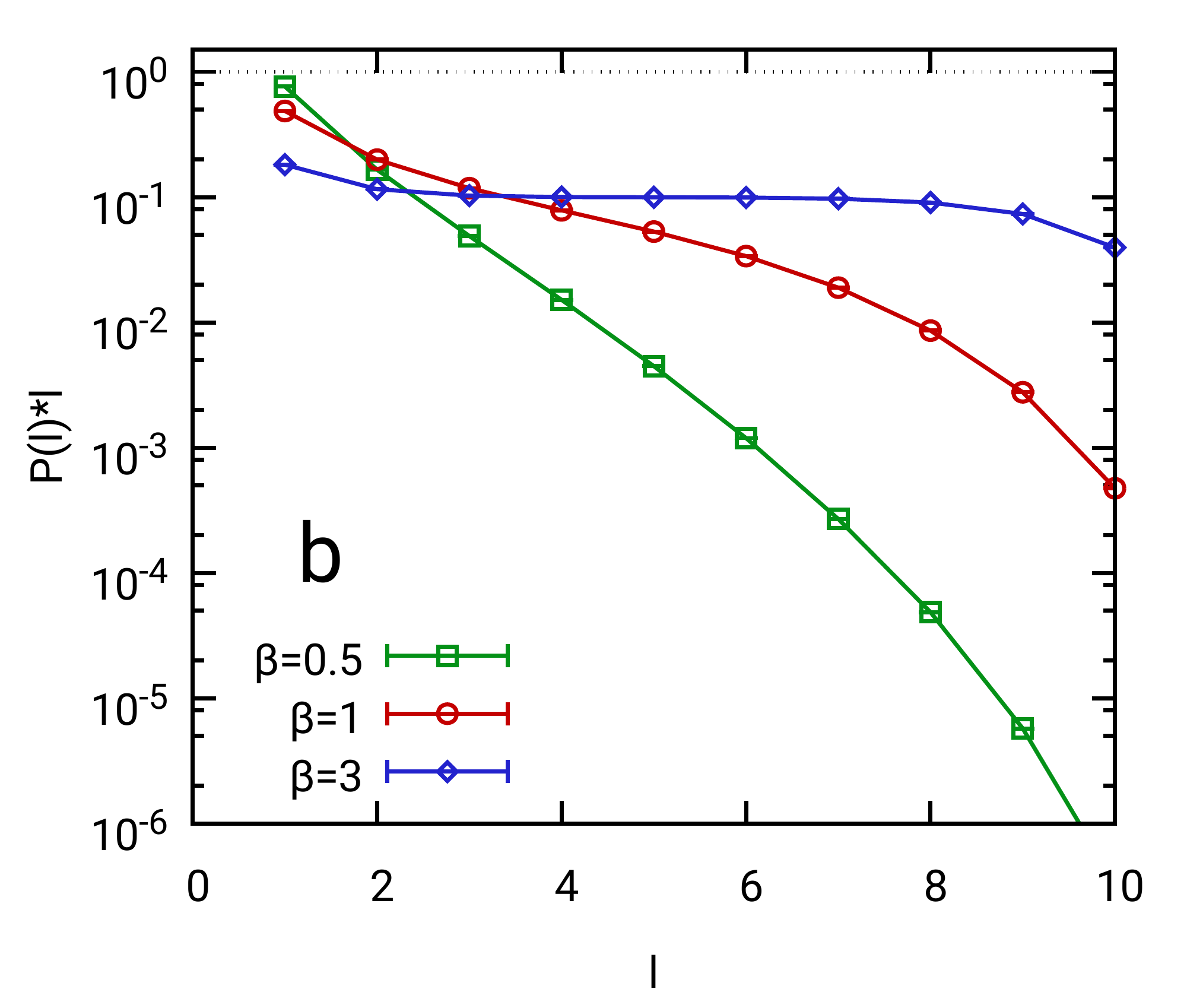}\\
\centering
\includegraphics[width=0.4\textwidth]{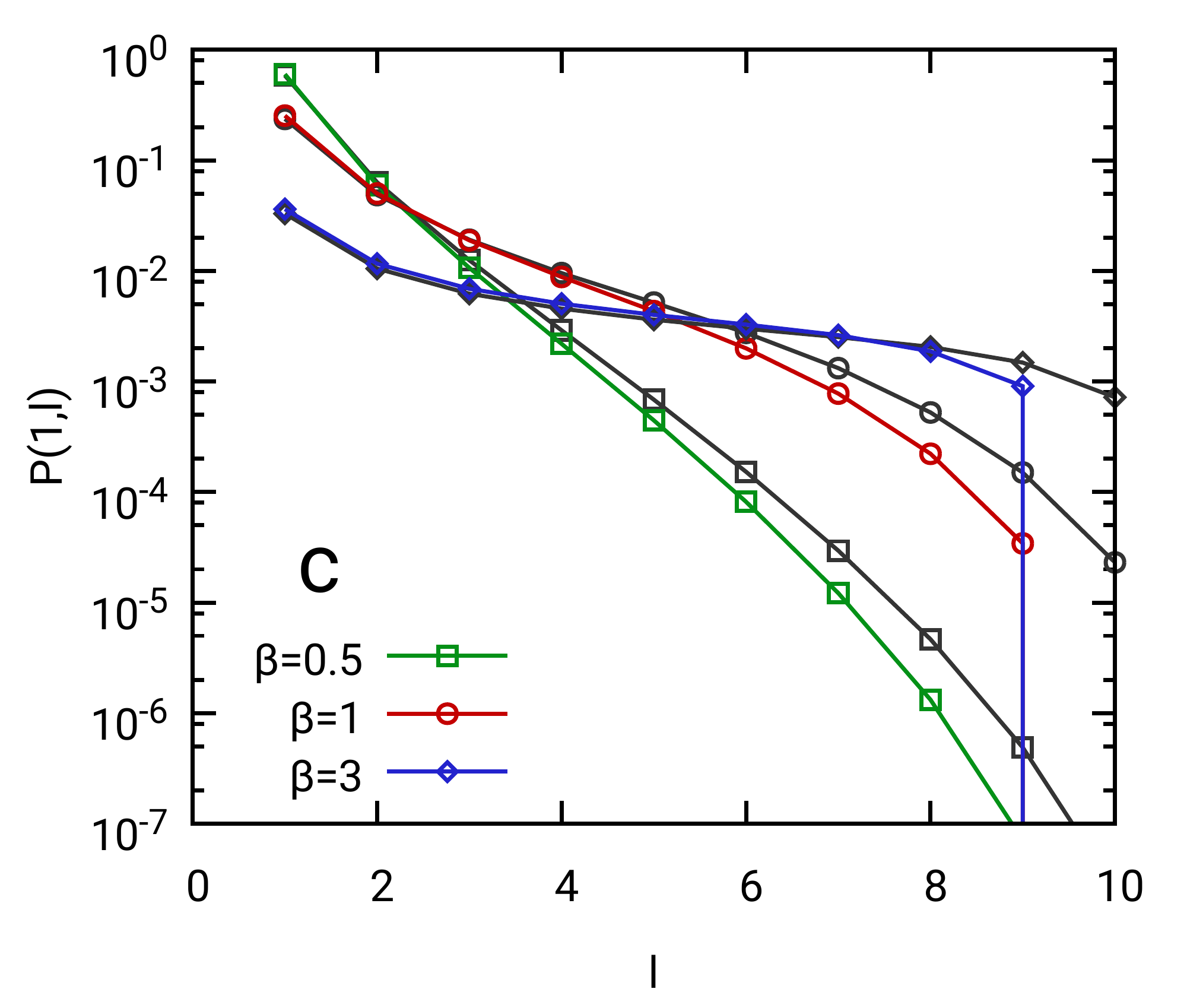}\includegraphics[width=0.4\textwidth]{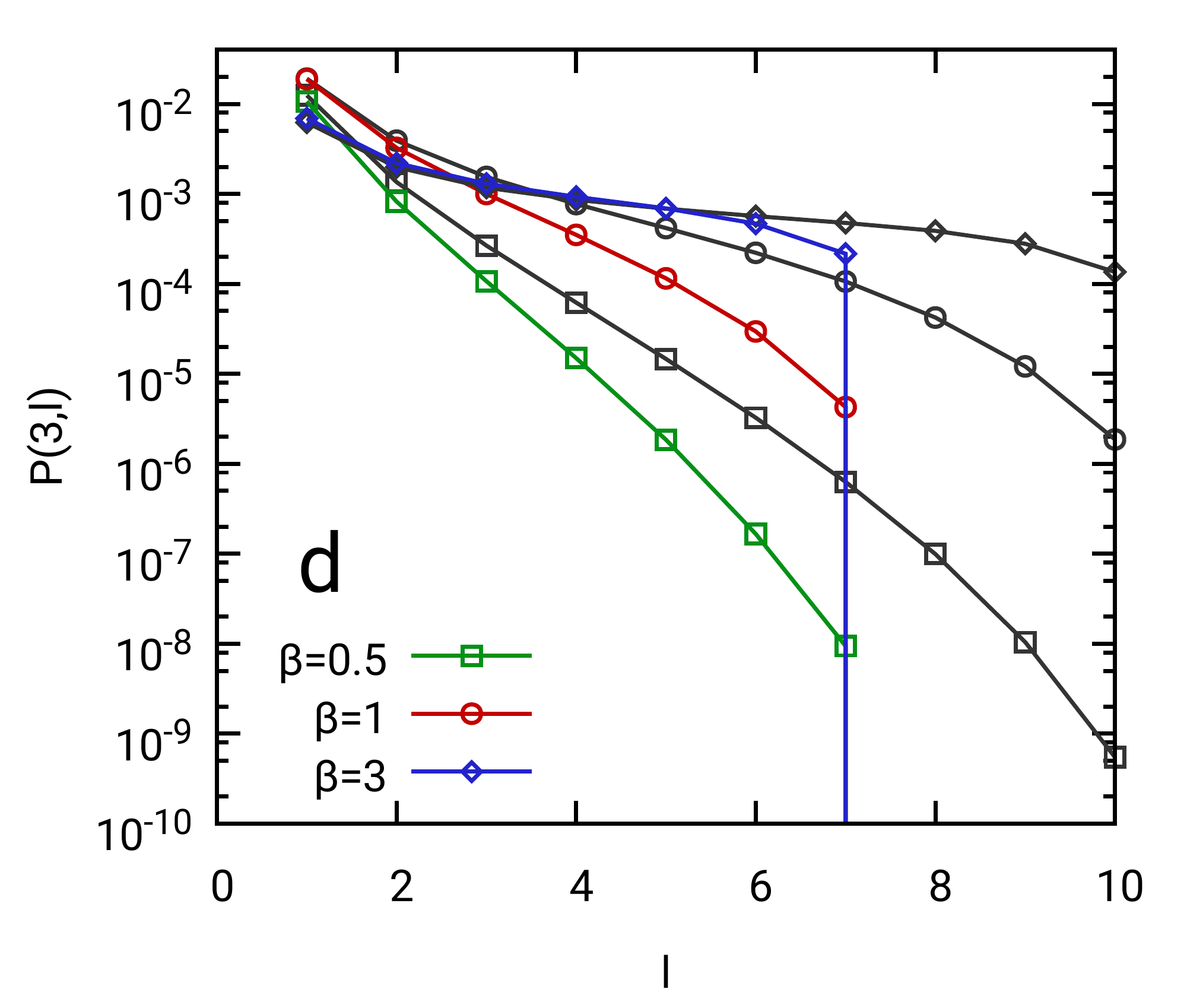}
\caption{\label{fig:column_trap}
PIMC results for the density and permutation cycle properties of $N=10$ electrons in a $2D$ harmonic trap with coupling strength $\lambda=0.5$ and $\beta=0.5,1,3$---Radial density distribution $n(r)$ for bosons and fermions in dependence of the distance to the center of the trap $r$ (panel a). Probability to find each particle in a permutation cycle of length $l$, $P(l)l$ for different inverse temperatures, (panel b), and segments of the permutation cycle pair distribution function $P(k,l)$ for $k=1,3$ (panels c,d), with the colored and dark grey symbols corresponding to our PIMC results and the uncorrelated analogue $P_\textnormal{u}(k,l)=P(l)P(k)$, respectively.
}
\end{figure*}

In Fig.~\ref{fig:column_trap} a), we show radial density profiles for $N=10$ electrons at a moderate temperature and weak coupling, $\beta=0.5$ and $\lambda=0.5$. The green curve corresponds to the bosonic density measured in the modified configuration space $Z'=Z_\textnormal{B}$ and the red curve to the corresponding fermioninc density profile that has been computed from Eq.~(\ref{eq:AS}). Evidently, bosons tend to cluster more around the centre of the trap, whereas the Pauli blocking acts as an effective repulsive force~\cite{CARSON199699}, which leads to a more pronounced spatial separation and pushes the electrons outwards. We again stress that both data sets have been obtained within the same PIMC simulation and the differences between bosons and fermions are purely a result of the cancellation of positive and negative terms, see Ref.~\onlinecite{doi:10.1002/ctpp.201800157} for an extensive topical discussion.

In panel b), we show the corresponding permutation-cycle probabilities $P(l)l$ for $\beta=0.5$ (green squares), $\beta=1$ (red circles), and $\beta=3$ (blue diamonds). Note that the average signs are given by $S=0.0615(1)$, $S=0.00022(4)$, and $S=-0.00003(3)$, respectively, which means that PIMC simulations are only feasible at $\beta=0.5$. In contrast, for the lowest considered temperature, the sign vanishes within the given statistical uncertainty.
For the largest temperature, we find a smooth decay with $l$, whereas for $\beta=3$ the probability to find a particle in an exchange-cycle of length $l$ is again almost constant, see also the description of Fig.~\ref{fig:Pl_T} for a more extensive discussion.


Finally, in panels c) and d), we show our PIMC results for the joint probability $P(1,l)$ and $P(3,l)$, and compare them to the corresponding uncorrelated function $P_\textnormal{u}(l,k)$ (dark grey symbols). Similar to the previously discussed case of the warm dense electron gas, we find that correlations between exchange-cycles are more pronounced for higher temperature, and increase for larger cycle-lengths $l$.
However, in contrast to the UEG, in the case of an inhomogeneous system in an external potential, any finite-size effects have real physical meaning and are not merely an artifact due to the inevitably limited finite simulation box within a PIMC simulation.
Therefore, approximation schemes that rely on the absence of correlations between different permutation-cycles within a given configuration as proposed in Ref.~\onlinecite{dubois_permutation} are not applicable in this case.

\section{Summary and discussion\label{sec:discusson}}

In summary, we have provided a detailed analysis of the permutation-cycle properties of path-integral Monte-Carlo simulations of correlated, quantum degenerate electrons. To verify our implementation, and to further analyse our results for the warm dense electron gas, we have begun our investigations by considering $N$ ideal spin-polarised fermions in periodic box. Even for this most simple case, we find an exponential decay in the permutation-cycle distribution function $P(l)$ only at relatively high temperatures, $\theta\gtrsim 1$. In addition, this quantity exhibits severe finite-size effects, which increase with the cycle length $l$. Similarly, the corresponding analysis of the permutation-cycle correlation-function $P(l,k)$ has revealed that cycles of different lengths are independent only for small $l$, whereas correlations increase for larger $l$, again due to the finite number of particles within the simulation. In contrast, permutation-cycles of noninteracting fermions are completely independent in the thermodynamic limit.

Extending our considerations to the warm dense electron gas, here, too, we have found that exchange-cycles are almost independent for small $l$, and deviations between $P(l,k)$ and $P_\textnormal{u}(k,l)=P(k)P(l)$ only start to become important for long cycles. Due to the striking similarity to the previously investigated ideal case, we have concluded that correlation effects of permutation-cycle properties within a PIMC simulation are dominated by finite-size effects and not, as it might have been expected, by the Coulomb interaction. 
In addition, we have also investigated the temperature-dependence of these properties, which, again, has revealed a remarkable behaviour: at low temperatures, where the system is most strongly correlated, the permutation properties of the electron gas most closely resemble the exact behavior of ideal electrons; at high temperatures, where correlation effects are expected to be less important, the ideal and interacting system exhibit more pronounced deviations. The explanation for this rather peculiar finding is the ground-state limit for the permutation-cycle distribution function, $P(l)l = 1/N$, which is equal both in the interacting and noninteracting cases.

In a nutshell, we have found that the permutation-cycle properties of interacting electrons are qualitatively very similar to the noninteracting case. In particular, cycles of different lengths are surprisingly uncorrelated, which can potentially be exploited to significantly simplify the fermionic configuration space~\cite{dubois_permutation}, and to extend fermionic PIMC simulations to lower temperatures and stronger quantum degeneracy. Moreover, the observed correlation effects in $P(l,k)$ for large $l$ and $k$ appear to be an artifact due to the finite number of electrons within a given simulation. Therefore, ignoring them might not necessarily be a shortcoming, but rather a handy way to mitigate finite-size effects~\cite{PhysRevB.78.125106,PhysRevB.94.035126}.

Finally, we have further extended our considerations to a finite system in an external harmonic potential. In this case, however, the observed finite-size and correlation effects in the permutation-cycle properties are not of an artificial nature, but instead represent real physical behaviour. Therefore, neglecting correlations between cycles is not appropriate, even for noninteracting fermions.

We expect our results to be of interest for the further development of PIMC simulations of fermions, which, in turn is of high importance for many fields like the simulation of ultracold atoms~\cite{RevModPhys.80.885,PhysRevLett.74.2288,Bernu1992}, quark-gluon plasmas~\cite{PhysRevC.87.035207,doi:10.1002/ctpp.201400056}, and warm dense matter~\cite{dornheim_review,10.1007/978-3-319-04912-0_5}.


 \section*{Acknowledgments}
 We thank B.~Alder for discussions on the topic presented in Ref.~\onlinecite{dubois_permutation}.
This work has been supported by the Deutsche Forschungsgemeinschaft via project BO1366/13 and by the Norddeutscher Verbund f\"ur Hoch- und H\"ochleistungsrechnen (HLRN) via grant shp00015 for CPU time.





\bibliographystyle{aps}

\bibliography{main.bib}

\begin{thebibliography}{10}
\providecommand{\url}[1]{\texttt{#1}}
\providecommand{\urlprefix}{URL }
\providecommand{\eprint}[2][]{\url{#2}}

\bibitem{berne_JCP_1982}
M.~F. Herman, E.~J. Bruskin, and B.~J. Berne, {On path integral Monte Carlo
  simulations}, J. Chem. Phys. \textbf{76}, 5150 (1982).

\bibitem{pollock_PRB_1984}
E.~L. Pollock and D.~M. Ceperley, {Simulation of quantum many-body systems by
  path-integral methods}, Phys. Rev. B \textbf{30}, 2555 (1984).

\bibitem{ceperley_path_1995}
D.~M. Ceperley, Path integrals in the theory of condensed helium, Reviews of
  Modern Physics \textbf{67}, 279 (1995).

\bibitem{sindzingre_path-integral_1989}
P.~Sindzingre, M.~L. Klein, and D.~M. Ceperley, {Path-integral {Monte} {Carlo}
  study of low-temperature $^4$He clusters}, Phys. Rev. Lett. \textbf{63}, 1601
  (1989).

\bibitem{kwon_local_2006}
Y.~Kwon, F.~Paesani, and K.~Whaley, Local superfluidity in inhomogeneous
  quantum fluids, Phys. Rev. B \textbf{74} (2006).

\bibitem{dornheim_superfluidity_2015}
T.~Dornheim, A.~Filinov, and M.~Bonitz, Superfluidity of strongly correlated
  bosons in two- and three-dimensional traps, Phys. Rev. B \textbf{91}, 054503
  (2015).

\bibitem{pilati_dilute_2010}
S.~Pilati, S.~Giorgini, M.~Modugno, and N.~Prokof'ev, Dilute {Bose} gas with
  correlated disorder: a path integral {Monte} {Carlo} study, New J. Phys.
  \textbf{12}, 073003 (2010).

\bibitem{noauthor_path-integral_2016}
H.~Saito, Path-{Integral} {Monte} {Carlo} {Study} on a {Droplet} of a {Dipolar}
  {Bose}–{Einstein} {Condensate} {Stabilized} by {Quantum} {Fluctuation}, J.
  Phys. Soc. Japan \textbf{85}, 053001 (2016).

\bibitem{boninsegni_worm_2006-1}
M.~Boninsegni, N.~V. Prokof’ev, and B.~V. Svistunov, Worm algorithm and
  diagrammatic {Monte} {Carlo}: {A} new approach to continuous-space path
  integral {Monte} {Carlo} simulations, Phys. Rev. E \textbf{74}, 036701
  (2006).

\bibitem{boninsegni_worm_2006}
M.~Boninsegni, N.~Prokof’ev, and B.~Svistunov, Worm {Algorithm} for
  {Continuous}-{Space} {Path} {Integral} {Monte}{\textasciitilde}{Carlo}
  {Simulations}, Phys. Rev. Lett. \textbf{96}, 070601 (2006).

\bibitem{loh_sign}
E.~Y. Loh, J.~E. Gubernatis, R.~T. Scalettar, S.~R. White, D.~J. Scalapino, and
  R.~L. Sugar, {Sign problem in the numerical simulation of many-electron
  systems}, Phys. Rev. B \textbf{41}, 9301 (1990).

\bibitem{doi:10.1063/1.4977920}
T.~Dornheim, S.~Groth, F.~D. Malone, T.~Schoof, T.~Sjostrom, W.~M.~C. Foulkes,
  and M.~Bonitz, {Ab initio quantum Monte Carlo simulation of the warm dense
  electron gas}, Phys. Plasmas \textbf{24}, 056303 (2017).

\bibitem{RevModPhys.80.885}
I.~Bloch, J.~Dalibard, and W.~Zwerger, {Many-body physics with ultracold
  gases}, Rev. Mod. Phys. \textbf{80}, 885 (2008).

\bibitem{PhysRevLett.74.2288}
M.~Boninsegni and D.~M. Ceperley, {Path Integral Monte Carlo Simulation of
  Isotopic Liquid Helium Mixtures}, Phys. Rev. Lett. \textbf{74}, 2288 (1995).

\bibitem{Bernu1992}
B.~Bernu, D.~Ceperley, and C.~Lhuillier, Specific heat and curie temperature
  evaluation on triangular lattice: Exchange frequencies for3he adsorbed on
  grafoil, evaluated by path integral techniques, J. low Temp. Phys.
  \textbf{89}, 589 (1992).

\bibitem{PhysRevB.72.035122}
A.~N. Rubtsov, V.~V. Savkin, and A.~I. Lichtenstein, {Continuous-time quantum
  Monte Carlo method for fermions}, Phys. Rev. B \textbf{72}, 035122 (2005).

\bibitem{0295-5075-90-1-10004}
E.~Kozik, K.~V. Houcke, E.~Gull, L.~Pollet, N.~Prokof'ev, B.~Svistunov, and
  M.~Troyer, {Diagrammatic Monte Carlo for correlated fermions}, EPL
  (Europhysics Letters) \textbf{90}, 10004 (2010).

\bibitem{Cheuk1260}
L.~W. Cheuk, M.~A. Nichols, K.~R. Lawrence, M.~Okan, H.~Zhang, E.~Khatami,
  N.~Trivedi, T.~Paiva, M.~Rigol, and M.~W. Zwierlein, {Observation of spatial
  charge and spin correlations in the 2D Fermi-Hubbard model}, Science
  \textbf{353}, 1260 (2016).

\bibitem{PhysRevC.87.035207}
V.~S. Filinov, Y.~B. Ivanov, V.~E. Fortov, M.~Bonitz, and P.~R. Levashov,
  {Color path-integral Monte-Carlo simulations of quark-gluon plasma:
  Thermodynamic and transport properties}, Phys. Rev. C \textbf{87}, 035207
  (2013).

\bibitem{doi:10.1002/ctpp.201400056}
V.~Filinov, M.~Bonitz, Y.~Ivanov, E.-M. Ilgenfritz, and V.~Fortov,
  {Thermodynamics of the Quark-Gluon Plasma at Finite Chemical Potential: Color
  Path Integral Monte Carlo Results}, Contrib. Plasma Phys. \textbf{55}, 203
  (2015).

\bibitem{PhysRevLett.120.115703}
M.~Sch\"ottler and R.~Redmer, {Ab Initio Calculation of the Miscibility Diagram
  for Hydrogen-Helium Mixtures}, Phys. Rev. Lett. \textbf{120}, 115703 (2018).

\bibitem{PhysRevLett.120.025701}
G.~Mazzola, R.~Helled, and S.~Sorella, {Phase Diagram of Hydrogen and a
  Hydrogen-Helium Mixture at Planetary Conditions by Quantum Monte Carlo
  Simulations}, Phys. Rev. Lett. \textbf{120}, 025701 (2018).

\bibitem{PhysRevB.75.024206}
J.~Vorberger, I.~Tamblyn, B.~Militzer, and S.~A. Bonev, Hydrogen-helium
  mixtures in the interiors of giant planets, Phys. Rev. B \textbf{75}, 024206
  (2007).

\bibitem{PhysRevB.84.224109}
S.~X. Hu, B.~Militzer, V.~N. Goncharov, and S.~Skupsky, First-principles
  equation-of-state table of deuterium for inertial confinement fusion
  applications, Phys. Rev. B \textbf{84}, 224109 (2011).

\bibitem{PhysRevE.90.033111}
S.~X. Hu, L.~A. Collins, V.~N. Goncharov, T.~R. Boehly, R.~Epstein, R.~L.
  McCrory, and S.~Skupsky, {First-principles opacity table of warm dense
  deuterium for inertial-confinement-fusion applications}, Phys. Rev. E
  \textbf{90}, 033111 (2014).

\bibitem{falk_2018}
K.~Falk, {Experimental methods for warm dense matter research}, High Power
  Laser Science and Engineering \textbf{6}, e59 (2018).

\bibitem{seitz1943physics}
F.~Seitz, \emph{The physics of metals}, Metallurgy and metallurgical
  engineering series  (McGraw-Hill 1943).

\bibitem{giuliani2005quantum}
G.~Giuliani and G.~Vignale, \emph{{Quantum Theory of the Electron Liquid}},
  Masters Series in Physics and Astronomy  (Cambridge University Press 2005).

\bibitem{graziani2014frontiers}
F.~Graziani, M.~Desjarlais, R.~Redmer, and S.~Trickey, \emph{{Frontiers and
  Challenges in Warm Dense Matter}}, Lecture Notes in Computational Science and
  Engineering  (Springer International Publishing 2014).

\bibitem{10.1007/978-3-319-04912-0_5}
E.~Brown, M.~A. Morales, C.~Pierleoni, and D.~Ceperley, {Quantum Monte Carlo
  Techniques and Applications for Warm Dense Matter}, in F.~Graziani, M.~P.
  Desjarlais, R.~Redmer, and S.~B. Trickey, eds., \emph{Frontiers and
  Challenges in Warm Dense Matter}  (Springer International Publishing, Cham
  2014), pp. 123--149.

\bibitem{dornheim_review}
T.~Dornheim, S.~Groth, and M.~Bonitz, The uniform electron gas at warm dense
  matter conditions, Physics Reports \textbf{744}, 1  (2018).

\bibitem{dornheim_NJP2015}
T.~Dornheim, S.~Groth, A.~Filinov, and M.~Bonitz, {Permutation blocking path
  integral Monte Carlo: a highly efficient approach to the simulation of
  strongly degenerate non-ideal fermions}, New J. Phys. \textbf{17}, 073017
  (2015).

\bibitem{doi:10.1063/1.4936145}
T.~Dornheim, T.~Schoof, S.~Groth, A.~Filinov, and M.~Bonitz, {Permutation
  blocking path integral Monte Carlo approach to the uniform electron gas at
  finite temperature}, J. Chem. Phys. \textbf{143}, 204101 (2015).

\bibitem{PhysRevLett.117.156403}
T.~Dornheim, S.~Groth, T.~Sjostrom, F.~D. Malone, W.~M.~C. Foulkes, and
  M.~Bonitz, {Ab Initio Quantum Monte Carlo Simulation of the Warm Dense
  Electron Gas in the Thermodynamic Limit}, Phys. Rev. Lett. \textbf{117},
  156403 (2016).

\bibitem{PhysRevLett.119.135001}
S.~Groth, T.~Dornheim, T.~Sjostrom, F.~D. Malone, W.~M.~C. Foulkes, and
  M.~Bonitz, {Ab initio Exchange-Correlation Free Energy of the Uniform
  Electron Gas at Warm Dense Matter Conditions}, Phys. Rev. Lett. \textbf{119},
  135001 (2017).

\bibitem{PhysRevB.89.245124}
N.~S. Blunt, T.~W. Rogers, J.~S. Spencer, and W.~M.~C. Foulkes, {Density-matrix
  quantum Monte Carlo method}, Phys. Rev. B \textbf{89}, 245124 (2014).

\bibitem{doi:10.1063/1.4927434}
F.~D. Malone, N.~S. Blunt, J.~J. Shepherd, D.~K.~K. Lee, J.~S. Spencer, and
  W.~M.~C. Foulkes, {Interaction picture density matrix quantum Monte Carlo},
  J. Chem. Phys. \textbf{143}, 044116 (2015).

\bibitem{PhysRevLett.117.115701}
F.~D. Malone, N.~S. Blunt, E.~W. Brown, D.~K.~K. Lee, J.~S. Spencer, W.~M.~C.
  Foulkes, and J.~J. Shepherd, {Accurate Exchange-Correlation Energies for the
  Warm Dense Electron Gas}, Phys. Rev. Lett. \textbf{117}, 115701 (2016).

\bibitem{PhysRevLett.115.050603}
N.~S. Blunt, A.~Alavi, and G.~H. Booth, {Krylov-Projected Quantum Monte Carlo
  Method}, Phys. Rev. Lett. \textbf{115}, 050603 (2015).

\bibitem{doi:10.1021/acs.jctc.8b00569}
Y.~Liu, M.~Cho, and B.~Rubenstein, {Ab Initio Finite Temperature Auxiliary
  Field Quantum Monte Carlo}, J. Chem. Theory Comput. \textbf{14}, 4722 (2018),
  pMID: 30102856.

\bibitem{PhysRevB.95.205109}
J.~Claes and B.~K. Clark, {Finite-temperature properties of strongly correlated
  systems via variational Monte Carlo}, Phys. Rev. B \textbf{95}, 205109
  (2017).

\bibitem{PhysRevLett.115.176403}
B.~Militzer and K.~P. Driver, {Development of Path Integral Monte Carlo
  Simulations with Localized Nodal Surfaces for Second-Row Elements}, Phys.
  Rev. Lett. \textbf{115}, 176403 (2015).

\bibitem{PhysRevE.91.033108}
V.~S. Filinov, V.~E. Fortov, M.~Bonitz, and Z.~Moldabekov, {Fermionic
  path-integral Monte Carlo results for the uniform electron gas at finite
  temperature}, Phys. Rev. E \textbf{91}, 033108 (2015).

\bibitem{universe4120133}
V.~Filinov and A.~Larkin, {Quantum Dynamics of Charged Fermions in the Wigner
  Formulation of Quantum Mechanics}, Universe \textbf{4} (2018).

\bibitem{PhysRevE.96.023203}
T.~Dornheim, S.~Groth, J.~Vorberger, and M.~Bonitz, {Permutation-blocking
  path-integral Monte Carlo approach to the static density response of the warm
  dense electron gas}, Phys. Rev. E \textbf{96}, 023203 (2017).

\bibitem{doi:10.1063/1.4999907}
S.~Groth, T.~Dornheim, and M.~Bonitz, {Configuration path integral Monte Carlo
  approach to the static density response of the warm dense electron gas}, J.
  Chem. Phys. \textbf{147}, 164108 (2017).

\bibitem{PhysRevLett.121.255001}
T.~Dornheim, S.~Groth, J.~Vorberger, and M.~Bonitz, {Ab initio Path Integral
  Monte Carlo Results for the Dynamic Structure Factor of Correlated Electrons:
  From the Electron Liquid to Warm Dense Matter}, Phys. Rev. Lett.
  \textbf{121}, 255001 (2018).

\bibitem{PhysRevLett.110.146405}
E.~W. Brown, B.~K. Clark, J.~L. DuBois, and D.~M. Ceperley, {Path-Integral
  Monte Carlo Simulation of the Warm Dense Homogeneous Electron Gas}, Phys.
  Rev. Lett. \textbf{110}, 146405 (2013).

\bibitem{PhysRevB.93.085102}
S.~Groth, T.~Schoof, T.~Dornheim, and M.~Bonitz, {Ab initio quantum Monte Carlo
  simulations of the uniform electron gas without fixed nodes}, Phys. Rev. B
  \textbf{93}, 085102 (2016).

\bibitem{PhysRevB.93.205134}
T.~Dornheim, S.~Groth, T.~Schoof, C.~Hann, and M.~Bonitz, {Ab initio quantum
  Monte Carlo simulations of the uniform electron gas without fixed nodes: The
  unpolarized case}, Phys. Rev. B \textbf{93}, 205134 (2016).

\bibitem{doi:10.1002/ctpp.201700096}
T.~Dornheim, S.~Groth, and M.~Bonitz, Ab initio results for the static
  structure factor of the warm dense electron gas, Contrib. Plasma Phys.
  \textbf{57}, 468 (2017).

\bibitem{doi:10.1002/ctpp.201600082}
S.~Groth, T.~Dornheim, and M.~Bonitz, {Free energy of the uniform electron gas:
  Testing analytical models against first-principles results}, Contrib. Plasma
  Phys. \textbf{57}, 137 (2017).

\bibitem{PhysRevA.48.4075}
A.~P. Lyubartsev and P.~N. Vorontsov-Velyaminov, {Path-integral Monte Carlo
  method in quantum statistics for a system of N identical fermions}, Phys.
  Rev. A \textbf{48}, 4075 (1993).

\bibitem{PhysRevE.80.066702}
M.~A. Voznesenskiy, P.~N. Vorontsov-Velyaminov, and A.~P. Lyubartsev,
  {Path-integral--expanded-ensemble Monte Carlo method in treatment of the sign
  problem for fermions}, Phys. Rev. E \textbf{80}, 066702 (2009).

\bibitem{dubois_permutation}
J.~L. DuBois, E.~W. Brown, and B.~J. Alder, {Overcoming the Fermion Sign
  Problem in Homogeneous Systems}, in E.~Schwegler, B.~M. Rubenstein, and S.~B.
  Libby, eds., \emph{Advances in the Computational Sciences---Symposium in
  Honor of Dr Berni Alder's 90th Birthday} (2017), pp. 184--192.

\bibitem{trotter_product_1959}
H.~F. Trotter, On the {Product} of {Semi}-{Groups} of {Operators},
  Proc.~Am.~Math.~Soc. \textbf{10}, 545 (1959).

\bibitem{de_raedt_applications_1983}
H.~De~Raedt and B.~De~Raedt, Applications of the generalized {Trotter} formula,
  Phys.~Rev.~A \textbf{28}, 3575 (1983).

\bibitem{chandler_exploiting_1981}
D.~Chandler and P.~Wolynes, Exploiting the isomorphism between quantum theory
  and classical statistical mechanics of polyatomic fluids, J.~Chem.~Phys.
  \textbf{74}, 4078 (1981).

\bibitem{yellow_book}
A.~Filinov and M.~Bonitz, {Classical and Quantum Monte Carlo}, in M.~Bonitz and
  D.~Semkat, eds., \emph{Introduction to Computational Methods in Many Body
  Physics}  (Rinton Press, Paramus, New Jersey (United States of America)
  2006), chapter~5, p. 237.

\bibitem{liu2013monte}
J.~Liu, \emph{{Monte Carlo Strategies in Scientific Computing}}, Springer
  Series in Statistics  (Springer New York 2013).

\bibitem{metropolis_equation_1953}
N.~Metropolis, A.~W. Rosenbluth, M.~N. Rosenbluth, A.~H. Teller, and E.~Teller,
  {Equation of {State} {Calculations} by {Fast} {Computing} {Machines}}, J.
  Chem. Phys. \textbf{21}, 1087 (1953).

\bibitem{PhysRevLett.96.105301}
M.~Boninsegni, N.~Prokof'ev, and B.~Svistunov, {Superglass Phase of
  $^{4}\mathrm{He}$}, Phys. Rev. Lett. \textbf{96}, 105301 (2006).

\bibitem{PhysRevLett.97.045301}
F.~Mezzacapo and M.~Boninsegni, {Superfluidity and Quantum Melting of
  $p\mathrm{\text{\ensuremath{-}}}{\mathrm{H}}_{2}$ Clusters}, Phys. Rev. Lett.
  \textbf{97}, 045301 (2006).

\bibitem{PhysRevLett.105.070401}
A.~Filinov, N.~V. Prokof'ev, and M.~Bonitz, {Berezinskii-Kosterlitz-Thouless
  Transition in Two-Dimensional Dipole Systems}, Phys. Rev. Lett. \textbf{105},
  070401 (2010).

\bibitem{PhysRevB.84.075130}
J.~B\"oning, A.~Filinov, and M.~Bonitz, Crystallization of an exciton
  superfluid, Phys. Rev. B \textbf{84}, 075130 (2011).

\bibitem{filinov_collective_2012}
A.~Filinov and M.~Bonitz, Collective and single-particle excitations in
  two-dimensional dipolar {Bose} gases, Phys. Rev. A \textbf{86} (2012).

\bibitem{filinov_correlation_2016}
A.~Filinov, Correlation effects and collective excitations in bosonic bilayers:
  {Role} of quantum statistics, superfluidity, and the dimerization transition,
  Phys. Rev. A \textbf{94}, 013603 (2016).

\bibitem{doi:10.1002/ctpp.201800157}
T.~Dornheim, S.~Groth, and M.~Bonitz, {Permutation blocking path integral Monte
  Carlo simulations of degenerate electrons at finite temperature}, Contrib.
  Plasma Phys. (in press) .

\bibitem{troyer_sign}
M.~Troyer and U.-J. Wiese, {Computational Complexity and Fundamental
  Limitations to Fermionic Quantum Monte Carlo Simulations}, Phys. Rev. Lett.
  \textbf{94}, 170201 (2005).

\bibitem{krauth_book_2006statistical}
W.~Krauth, \emph{{Statistical Mechanics: Algorithms and Computations}}, Oxford
  Master Series in Physics  (Oxford University Press, UK 2006).

\bibitem{Gill_UEGREVIEW_2011}
P.~M.~W. Gill and P.-F. Loos, Uniform electron gases, Theoretical Chemistry
  Accounts \textbf{131}, 1069 (2011).

\bibitem{fraser_PRB}
L.~M. Fraser, W.~M.~C. Foulkes, G.~Rajagopal, R.~J. Needs, S.~D. Kenny, and
  A.~J. Williamson, {Finite-size effects and Coulomb interactions in quantum
  Monte Carlo calculations for homogeneous systems with periodic boundary
  conditions}, Phys. Rev. B \textbf{53}, 1814 (1996).

\bibitem{Yakub_JCP}
E.~Yakub and C.~Ronchi, {An efficient method for computation of long-ranged
  Coulomb forces in computer simulation of ionic fluids}, J. Chem. Phys.
  \textbf{119}, 11556 (2003).

\bibitem{Yakub2005}
E.~Yakub and C.~Ronchi, A new method for computation of long ranged coulomb
  forces in computer simulation of disordered systems, Journal of Low
  Temperature Physics \textbf{139}, 633 (2005).

\bibitem{reimann_RevModPhys2002}
S.~M. Reimann and M.~Manninen, {Electronic structure of quantum dots}, Rev.
  Mod. Phys. \textbf{74}, 1283 (2002).

\bibitem{dornheim_CPP2016}
T.~Dornheim, H.~Thomsen, P.~Ludwig, A.~Filinov, and M.~Bonitz, {Analyzing
  Quantum Correlations Made Simple}, Contrib. Plasma Phys. \textbf{56}, 371
  (2016).

\bibitem{kylanpaa_PhysRevB2017}
I.~Kyl\"anp\"a\"a and E.~R\"as\"anen, {Path integral Monte Carlo benchmarks for
  two-dimensional quantum dots}, Phys. Rev. B \textbf{96}, 205445 (2017).

\bibitem{egger_PRL1998}
C.~H. Mak, R.~Egger, and H.~Weber-Gottschick, {Multilevel Blocking Approach to
  the Fermion Sign Problem in Path-Integral Monte Carlo Simulations}, Phys.
  Rev. Lett. \textbf{81}, 4533 (1998).

\bibitem{schoof_CPP2011}
T.~Schoof, M.~Bonitz, A.~Filinov, D.~Hochstuhl, and J.~Dufty, {Configuration
  Path Integral Monte Carlo}, Contrib. Plasma Phys. \textbf{51}, 687 (2011).

\bibitem{CARSON199699}
C.~Carson, {The peculiar notion of exchange forces-II: From nuclear forces to
  QED, 1929-1950}, Studies in History and Philosophy of Science Part B: Studies
  in History and Philosophy of Modern Physics \textbf{27}, 99  (1996).

\bibitem{PhysRevB.78.125106}
N.~D. Drummond, R.~J. Needs, A.~Sorouri, and W.~M.~C. Foulkes, {Finite-size
  errors in continuum quantum Monte Carlo calculations}, Phys. Rev. B
  \textbf{78}, 125106 (2008).

\bibitem{PhysRevB.94.035126}
M.~Holzmann, R.~C. Clay, M.~A. Morales, N.~M. Tubman, D.~M. Ceperley, and
  C.~Pierleoni, {Theory of finite size effects for electronic quantum Monte
  Carlo calculations of liquids and solids}, Phys. Rev. B \textbf{94}, 035126
  (2016).

\end{thebibliography}







\end{document}